\documentclass[twocolumn,superscriptaddress]{revtex4}
\usepackage[utf8]{inputenc}
\usepackage{xcolor}
\usepackage{graphics,graphicx,epsfig}
\usepackage{amssymb,amsfonts,amsmath}
\usepackage{ifthen}	
\usepackage{hyperref}	
\newcommand{\EQ}{\begin{equation}}
\newcommand{\EE}{\end{equation}}
\newcommand{\EQA}{\begin{eqnarray}}
\newcommand{\EEA}{\end{eqnarray}}
\newcommand{\D}{{\mathcal{D}}}
\newcommand{\G}{{\mathcal{G}}}
\newcommand{\Q}{{\mathcal{Q}}}

\newcommand{\gen}{{\text {gen}}}
\newcommand{\post}{{\text {post}}}

\newcommand{\x}{{\bf{x}}}
\newcommand{\seemethods}{{\text {see Methods}}}
\newcommand{\beqn}{\begin{eqnarray}}
\newcommand{\eeqn}{\end{eqnarray}}
\newcommand{\beq}{\begin{equation}}
\newcommand{\eeq}{\end{equation}}

\usepackage{pdfpages}
\begin{document}

\title{Deep generative selection models of T and B cell receptor repertoires with soNNia}
\author{Giulio Isacchini}
\address{Max Planck Institute for Dynamics and Self-organization, Am Fa\ss berg 17, 37077 G\"ottingen, Germany}
\address{Laboratoire de physique de l'\'ecole normale sup\'erieure (PSL University), CNRS, Sorbonne Universit\'e, and Universit\'e de Paris, 75005 Paris, France}
\author{Aleksandra M. Walczak}
\thanks{These authors contributed equally.\\ Correspondence should be addressed to: aleksandra.walczak@phys.ens.fr, thierry.mora@phys.ens.fr, and armita@uw.edu. }
\address{Laboratoire de physique de l'\'ecole normale sup\'erieure (PSL University), CNRS, Sorbonne Universit\'e, and Universit\'e de Paris, 75005 Paris, France}
\author{Thierry Mora}
\thanks{These authors contributed equally.\\ Correspondence should be addressed to: aleksandra.walczak@phys.ens.fr, thierry.mora@phys.ens.fr, and armita@uw.edu. }
\address{Laboratoire de physique de l'\'ecole normale sup\'erieure (PSL University), CNRS, Sorbonne Universit\'e, and Universit\'e de Paris, 75005 Paris, France}
\author{Armita Nourmohammad}
\thanks{These authors contributed equally.\\ Correspondence should be addressed to: aleksandra.walczak@phys.ens.fr, thierry.mora@phys.ens.fr, and armita@uw.edu. }
\address{Max Planck Institute for Dynamics and Self-organization, Am Fa\ss berg 17, 37077 G\"ottingen, Germany}
\address{Department of Physics, University of Washington, 3910 15th Ave Northeast, Seattle, WA 98195, USA}
\address{Fred Hutchinson Cancer Research Center, 1100 Fairview ave N, Seattle, WA 98109, USA}

\date{\today} 
\begin{abstract}
{
Subclasses of lymphocytes carry different functional roles to work
together and produce an immune response and lasting
immunity. Additionally to these functional roles, T and B-cell
lymphocytes rely on the diversity of their receptor chains to
recognize different pathogens. The lymphocyte subclasses emerge from
common ancestors generated with the same diversity of receptors during
selection processes. Here we leverage biophysical models of receptor
generation with machine learning models of selection to identify
specific sequence features characteristic of functional lymphocyte
repertoires and subrepertoires. Specifically, using only repertoire level sequence information, we classify CD4$^+$ and CD8$^+$ T-cells, find correlations between receptor chains arising during selection, and identify T-cell subsets that are targets of pathogenic epitopes. We also show examples of when simple linear classifiers do as well as more complex machine learning methods.  }
\end{abstract}

\maketitle

\section*{Introduction}
The adaptive immune system in vertebrates consists of highly diverse B- and T-cells whose unique receptors mount specific responses against a multitude of pathogens. These diverse receptors are generated through genomic rearrangement and sequence insertions and deletions, a process known as V(D)J recombination~\cite{Tonegawa1983,Davis1988}. Recognition of a pathogen by a T- or B-cell receptor is mediated through molecular interactions between an immune receptor protein and a pathogenic epitope. T-cell receptor proteins interact with short protein fragments (peptide antigens) from the pathogen that are presented by specialized pathogen presenting Major Histocompatibility Complexes (MHC) on cell surface. B-cell receptors interact directly with epitopes on  pathogenic surfaces. Upon an infection, cells carrying those specific receptors that recognize the infecting pathogen become activated and proliferate to  control and neutralize the infection. A fraction of these selected responding cells later contribute to the  memory repertoire that reacts more readily in future encounters. Unsorted immune receptors sampled  from an individual reflect  both the history of infections and the ongoing  responses to infecting pathogens.

\begin{figure*}[t!]
\begin{center}
\includegraphics[width=0.9\linewidth]{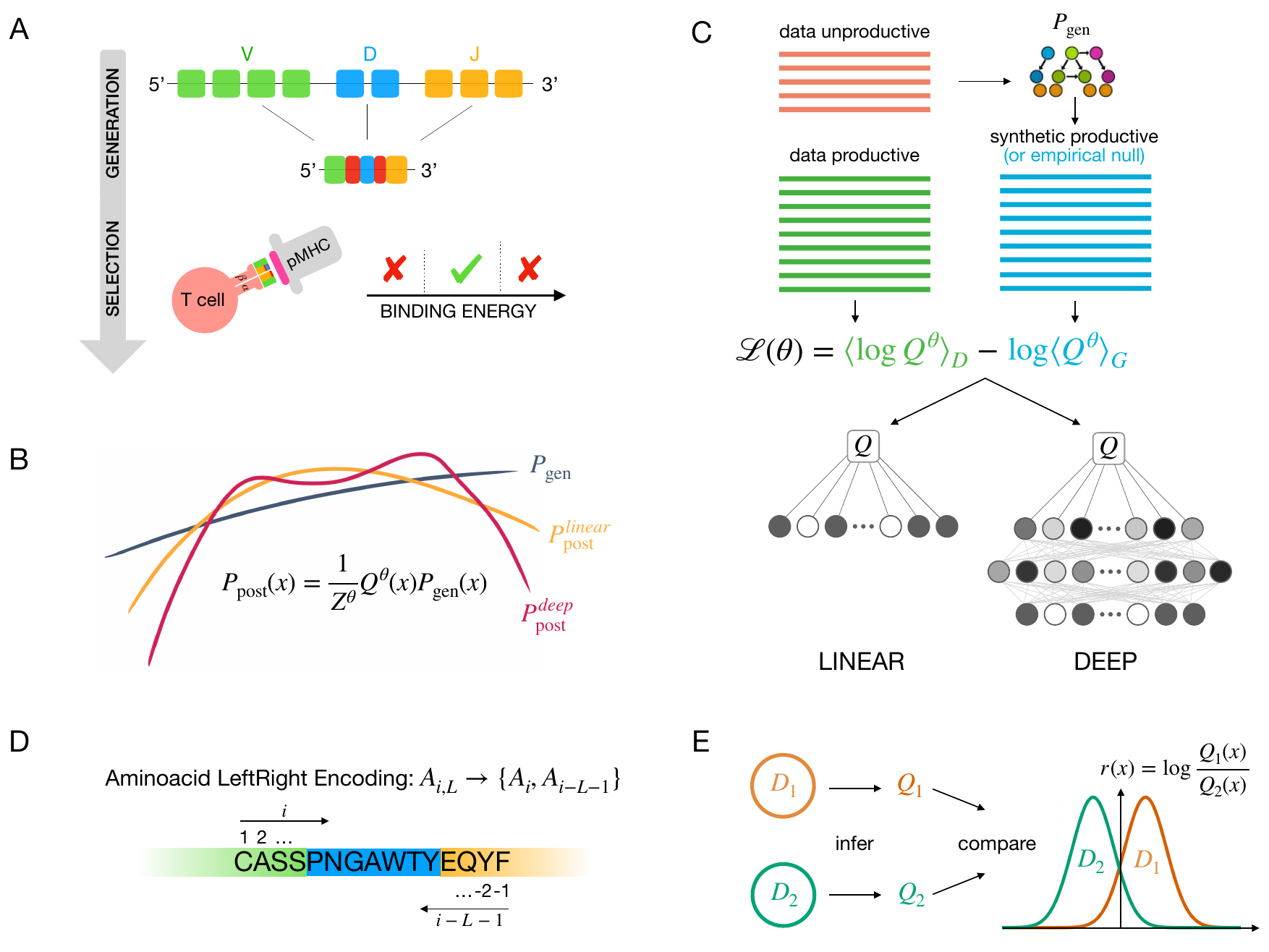}
\caption{{\bf Inference of functional selection models for immune receptor repertoires.}  {\bf (A) } T cell receptor $\alpha$ and $\beta$ chains are stochastically rearranged through a process called V(D)J recombination. Successfully rearranged receptors undergo selection for binding to self-pMHCs. Receptors that bind too weakly or too strongly are rejected, while intermediately binding ones exit the thymus and enter peripheral circulation. Development of B-cell receptors follows similar stages of stochastic recombination and selection.{\bf (B)}  We model these two processes independently. The statistics of the V(D)J recombination process described by the probability of generating a given receptor sequence $\sigma$, $P_{\gen}(\sigma)$, are inferred using the IGOR software \cite{Marcou2018}. $P_{\gen}(\sigma)$ acts as a baseline for the selection model. We then infer selection factors $\mathcal{Q}$, which act as weights that modulate the initial distribution $P_{\gen}(\sigma)$. We infer two types of selection weights: linear in log space (using the SONIA software \cite{Sethna2020}) and non-linear weights using a deep neural network, in the soNNia software presented here. Non-linear selection weights are more flexible than linear ones. {\bf (C)} Pipeline of the algorithm: $P_{\gen}$ is inferred from unproductive sequences using IGOR. Selection factors for both the linear and non-linear models are inferred from productive sequences by maximizing their log-likelihood $\mathcal{L}$, which involves a normalization term calculated by sampling unselected sequences generated by the OLGA software \cite{Sethna2019}. {\bf(D)} In both selection models the amino acid composition of the CDR3 is encoded by its relative distance from the left and right borders (left-right encoding). {\bf(E)} After inferring repertoire specific selection factors, repertoires are compared by computing e.g. log likelihood ratios $r(x)$.
\label{emersonfig}
}
\end{center}
\end{figure*}

Before entering the periphery where their role is to recognize foreign antigens, the generated receptors undergo a two-fold selection process based on their potential to bind to the organism's own self-proteins. On one-hand, they are tested to not be strongly self-reactive (Fig.~\ref{emersonfig}~A). On the other hand, they must be able to bind to some of the presented molecules to assure minimal binding capabilities. This pathogen-unspecific selection, known as thymic selection for T-cells~\cite{Yates2014} and the process of central tolerance in B-cells~\cite{Nemazee2017}, can prohibit over 90\% of generated receptors from entering the periphery~\cite{murphy2008janeway,Yates2014,Klein2014}.

Additionally to receptor diversity, T and B cell-subtypes are specialized to perform different functions. B- and T-cells in the adaptive immune system are differentiated from a common cell-type, known as lymphoid progenitor. T-cells differentiate into cell subtypes identified by their surface markers, including helper T-cells (CD4$^+$), killer T-cells (CD8$^+$)~\cite{Yates2014}, and regulatory T-cells or T-regs (CD4$^+$\,FOXP3$^+$)~\cite{Wing2010}, each of which can  be found in the non-antigen primed naive or memory compartment. The memory compartment can be further divided into subtypes, such as effector, central or stem cell-like memory cells, characterized by different lifetimes and roles. B-cells develop into, among other subtypes, plasmablasts and plasma cells, which are antibody factories, and memory cells that can be used against future infections. 
These cell subtypes perform distinct functions, react with different targets, and hence, experience different selection pressures. Here, we ask whether these different functions and selection pressures are reflected in their receptors' sequence compositions.

Recent progress in high-throughput immune repertoire sequencing (RepSeq) both for single-chain~\cite{Hou2016,Georgiou2014,Bolotin2015a,Mcdaniel2016} and paired-chain \cite{DeKosky2013a, Turchaninova2013c, Mcdaniel2016,Dekosky2014} B- and T-cell receptor has brought significant insight into the composition of immune repertoires. Based on such data, statistical inference techniques have been developed to infer biophysically informed sequence-based models for the underlying processes involved in generation and selection of immune receptors~\cite{Marcou2018,Ndifon2012,Ralph2016,Sethna2019,Sethna2020, Munshaw2010}. 
Machine learning techniques have also been used to infer deep generative models to characterize the T-cell repertoire composition as a whole~\cite{Davidsen2018}, as well as discriminate between public and private B-cell clones based on Complementarity Determining Region 3 (CDR3) sequence~\cite{Greiff2017a, Miho2019}. While biophysically informed models can still match and even outperform machine-learning techniques (see e.g.~\cite{Isacchini2020a}), deep learning models can be extremely powerful in describing functional subsets of immune repertoires, for which we lack a full biophysical understanding of the selection process. 

Here, we introduce a framework that uses the strengths of both biophysical models and machine learning approaches to characterize  signatures of differential selection acting on receptor sequences  from subsets associated with specific function. Specifically,  we leverage  biophysical tools to model what we know (e.g. receptor generation) and exploit the powerful machinery of deep neural networks (DNN) to model what we do not know (e.g. functional selection). Using the non-linear and flexible structure of the deep neural networks, we characterize the sequence properties that encode selection of the specificity of the combined chains during receptor maturation in $\alpha$ and$ \beta$ chains in T-cells, and  heavy and light ($\kappa$  and $\lambda$) chains in B-cells. We identify  informative sequence features that differentiate CD4$^+$ helper T-cells,  CD8$^+$ killer T-cells and regulatory T-cells. Finally, we demonstrate that that biophysical selection models can be used as simple classifiers to successfully identify T-cells specific to distinct targets of pathogenic epitopes--- a problem that is of significant interest for clinical applications \cite{Glanville2017,Shugay2018,Jokinen2019a,Gielis2019,Dash2017}.

\section*{Results}

\subsection*{Neural network models of TCR and BCR selection}
Previous work has inferred biophysically informed models of  V(D)J recombination underlying the generation of TCRs and BCRs~\cite{Murugan2012, Marcou2018}.  {In brief, these models are parametrized according to the probabilities by which different V-, D-, J- genes are used and base pairs are inserted in or deleted from the CDR3 junctions to generate a receptor sequence.} We infer the parameters of these models using the IGoR software~\cite{Marcou2018} from unproductive receptor sequences, which are generated, but due to a frameshift or insertion of stop codons are not expressed, and hence, are not subject to functional selection. The inferred models are used to characterize the generation probability of a receptor sequence $P_{\gen}$, and to synthetically generate an ensemble of pre-selection receptors~\cite{Sethna2019}. 
 These generated receptors define a baseline $\G$ for statistics of repertoires prior to any functional selection. 

{The amino acid sequence of an immune receptor protein determines its function.}
 To identify sequence properties that are linked to function, we compare the statistics of sequence features $f$ (e.g. V-, J- gene usage and CDR3 amino acid composition) present in a given B-  or T- cell functional repertoire to the expected baseline of receptor generation (Fig.~\ref{emersonfig} C). To do so, we encode a receptor sequence $\sigma$ as a binary vector $\x$ whose elements $x_f\in \{0,1\}$  specify whether the feature $f$ is present in a sequence  $\sigma$. The probability $P_{\rm post}^\theta(\x)$ for a given receptor  $\x$ to belong to a functional repertoire is described by modulating the receptor's generation probability $P_{\rm gen}(x)$ by a selection factor $\mathcal{Q}^\theta (\x)$,
\begin{equation}
P_{\rm post}^\theta(\x)=P_{\rm gen}(\x)\mathcal{Q}^\theta (\x) \equiv \frac{1}{ Z^\theta} P_\gen(\x) \,Q^\theta(\x),
\label{eq:model}
\end{equation}
where $\theta$ denotes the parameters of the selection model and $Z^\theta$ ensures normalization of $P_{\rm post}^\theta$. Previous work~\cite{Elhanati2014,Elhanati2015,Sethna2020} inferred selection models for functional repertoires by assuming a multiplicative form of selection $Q^\theta (\x)=\exp(\sum_{f}\theta^f x_f)$, where feature-specific factors $\theta^f$ contribute independently to selection. We refer to these models as linear SONIA (Fig.~\ref{emersonfig}B).  Selection can in general be a highly complex and non-linear function of the underlying sequence features. Here, we introduce soNNia,  a method to infer generic non-linear selection functions, using deep neural networks  (DNN). To infer a selection model that best describes sequence determinants of function in a data sample  $\D$, soNNia maximizes the mean log-likelihood of the data $\mathcal{L}(\theta) =\langle \log P_{\rm post}^\theta \rangle_\D$, where the probability  $P_{\rm post}^\theta$ is defined by Eq.~(\ref{eq:model}), and $\langle \cdot\rangle_\D$ denotes  expectation over the set of sequences $\D$. This likelihood can be rewritten as (\seemethods), 
 \begin{equation}
\begin{split}
\mathcal{L}(\theta) & =\langle \log P_{\rm post}^\theta \rangle_\D  =  \langle \log Q^\theta \rangle_\D - \log \langle Q^\theta
 \rangle_{\G}+{\rm const},
 \end{split}
\label{eq:likelihood}
 \end{equation}
 {where $\langle \cdot\rangle_{\G}$ is the expectation over the ensemble of sequences $\G$ that reflect the baseline.  This baseline set is often generated by sampling from a previously inferred generation model $P_\text{gen}$, using the  IGoR software~\cite{Marcou2018}.} Note that this expression becomes exact as the number of generated sequences approaches infinity.

In soNNia we divide the sequence features $f$ into three categories: (i) (V,J) usage, (ii) CDR3 length, and (iii) CDR3 amino acid composition encoded by a $20\times 50$ binary matrix that specifies the identity of an amino acid and its relative position within a 25 amino acid range from both the 5' and the 3' ends of the CDR3, equivalent to the left-right encoding of the SONIA model~\cite{Sethna2020} (Fig.~\ref{emersonfig}D). Input from each of the three categories are first propagated through their own network. Outputs from these three networks are then combined and transformed through a dense layer. This choice of architecture reduces the number of parameters in the DNN and makes the contributions of the three categories (which have different dimensions) comparable; \seemethods\  and  Figs.~S1,S3 and S4 for  details on the architecture of the DNN.

The baseline ensemble $\G$, which we have described as being generated from the $P_{\rm gen}$ model (Fig.~\ref{emersonfig} C), can in principle be replaced by any dataset, including empirical ones, at no additional computational cost, {for selection inference with soNNia.}   {This is especially useful when the goal is to only compare the selection models associated with different sub-repertoires with distinct functions. } We will use this feature of soNNia to learn selection coefficients of subsets relative to an {empirically constructed } generic functional repertoire. In that case, the inferred selection factors $Q$ only reflect differential selection relative to the generic baseline. {Importantly, this approach enables us to infer differential selection  without having to infer a common underlying generation model $P_{\rm gen}$ for the sub-repertoires.} Once two soNNia models have been learned from two distinct datasets, their statistics may be compared by computing a sequence-dependence log-likelihood ratio $r(x)=\log Q_1(x)/Q_2(x)$ predicting the preference of a sequence for a subset over the other. This log-likelihood ratio can be used as a functional classifier for receptor repertoires  (Fig.~\ref{emersonfig} E).

{
 
\subsection*{Deep non-linear selection model best describes functional TCR repertoire} 
\begin{figure*}[t!]
\includegraphics[width=\linewidth]{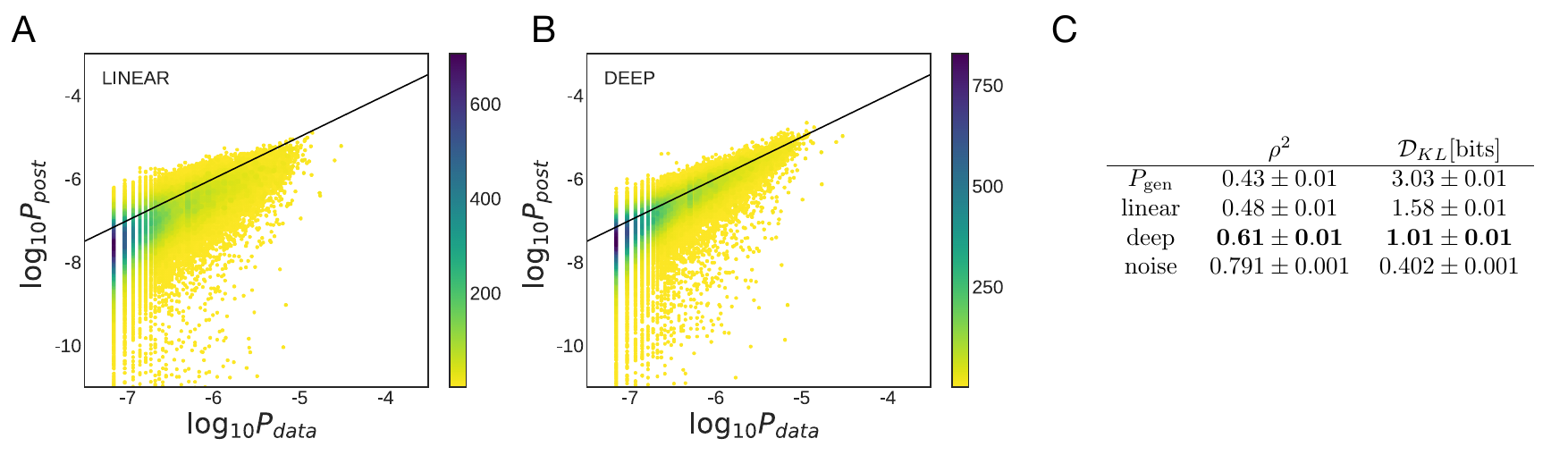}
\caption{
{\bf Performance of selection models on TCR repertoires.} Scatter plot of observed frequency, $P_{\rm data}$, versus predicted probability $P_{\rm post}$ for {\bf(A)} linear SONIA and {\bf(B)} deep neural network soNNia models trained on the TCR$\beta$ repertoires of 743 individuals from ref.~\cite{Emerson2017}. {The baseline is formed by sampling $10^7$ sequences from the $P_{\rm gen}$ model, learned from the nonproductive sequences of the same dataset (\seemethods).} Color indicates number of sequences.
 {\bf (C)} The soNNia model performs significantly better, as quantified by both the Kullback-Leibler divergence $\mathcal{D}_{\rm KL}$ (Methods) and the Pearson correlation coefficient $\rho^2$, {without overfitting (see Fig.~S2)}.
\label{emersonfigB}}
\end{figure*}

First, we systematically compare  the  accuracy of the (non-linear) soNNia model with linear SONIA~\cite{Sethna2020} (Fig.~\ref{emersonfig}~B) by inferring selection on TCR$\beta$ repertoires from a large cohort of  743 individuals from ref.~\cite{Emerson2017}.  
Our goal is to characterize selection on functional receptors irrespective of their phenotype. To avoid biases caused by expansions of particular receptors in different individuals, we pool the {\it unique} nucleotide sequences of receptors from all individuals and construct a universal donor. Multiplicity of an amino-acid sequence in this universal donor indicates the number of independent recombination events that have led to that receptor (in different individuals, or in the same individual by convergent recombination).

{We randomly split the pooled dataset into a training and a test set of equal sizes and trained both a SONIA and a soNNia selection model on the training set (Methods and Fig.~\ref{emersonfig}~C).  Our inference is highly stable, and the selection models are reproducible when trained on subsets of the training data ({\seemethods} and Fig. S2).}

To assess the  performance of our selection models, we compared their inferred probabilities $P_\post (\x)$ with the observed frequencies of the receptor sequences $P_{data}(\x)$ in the test set (Fig.~\ref{emersonfigB}A and B). Prediction accuracy can be quantified through the Pearson correlation between the two log-frequencies, or through their Kullback-Leibler divergence $\mathcal{D}_{\rm KL}(P_{\rm data}|P_{\rm post})$ (Methods and Fig.~\ref{emersonfigB}C). A smaller Kullback-Leibler divergence indicates a higher accuracy of the inferred model in predicting the data. The estimated accuracy of an inferred model is limited by the correlation between the test and the training set, which provides a lower bound on the Kullback-Leibler divergence $\mathcal{D}_{KL}\simeq 0.4$ bits, and an upper bound on the Pearson correlation $\rho^2\simeq0.8$.

We observe a substantial improvement of selection inference for the generalized selection model soNNia with $\mathcal{D}_{KL}\simeq 1.0$  bits (and Pearson correlation $\rho^2\simeq0.61$) compared to the linear SONIA model with  $\mathcal{D}_{KL}\simeq 1.6$ bits (and Pearson correlation $\rho^2\simeq0.48$); see Fig.~\ref{emersonfigB}.  Both models show a strong effect of selection, reducing the $\mathcal{D}_{KL}$ from $3.03$ bits (and increasing the correlation $\rho^2$ from $0.43$) for the comparison of data to the $P_{\rm gen}$ model alone (Fig.~\ref{emersonfigB}). This result highlights the role of complex nonlinear selection factors acting on receptor features that shape a functional T-cell repertoire.  The features that are still inaccessible to the soNNia selection factors are likely due to the sampling of rare features, {individual history of pathogenic exposures, or HLA differences among individuals.}

\label{sec_emerson}

\subsection*{Intra- and inter-chain interactions in TCRs and BCRs} 
T-cell receptors are disulfide-linked membrane-bound proteins made of variable  $\alpha$ and $\beta$ chains, and expressed as part of a complex that interact with pathogens. Similarly, B-cell receptors and antibodies are made up of a heavy and two major groups ($\kappa$ and $\lambda$) of light chains. Previous work has identified low but consistent correlations  between features of  $\alpha\beta$ chain pairs in T-cell receptors, with the largest contributions between $V_\alpha,V_\beta$ and $J_\alpha,V_\beta$~\cite{Grigaityte213462,Dupic2019,Carter2019,Tanno2020,Shcherbinin2020}. In B-cells, preferences for receptor features within heavy and light chains have been studied separately~\cite{Larimore2012,Glanville2009a} but inter-chain correlations have not been systematically investigated.

We first aimed to quantify dependencies between chains by re-analyzing recently published single-cell datasets: {TCR $\alpha\beta$ pairs of unfractionated repertoires from ref.~\cite{Tanno2020} and BCR of naive cells from ref.~\cite{DeKoskyE2636} (Methods)}. The blue bars of Fig.~\ref{fig:3} show the mutual information between the V and J choices and CDR3 length of each chain, for TCR $\alpha\beta$ (Fig.~\ref{fig:3}A), Ig H$\lambda$ (Fig.~\ref{fig:3}B), and Ig H$\kappa$ (Fig.~\ref{fig:3}C) repertoires. Mutual information is a non-parametric measure of correlation between pairs of variables (\seemethods).

\begin{figure*}[t!]
\begin{center}
\includegraphics[width=0.8\linewidth]{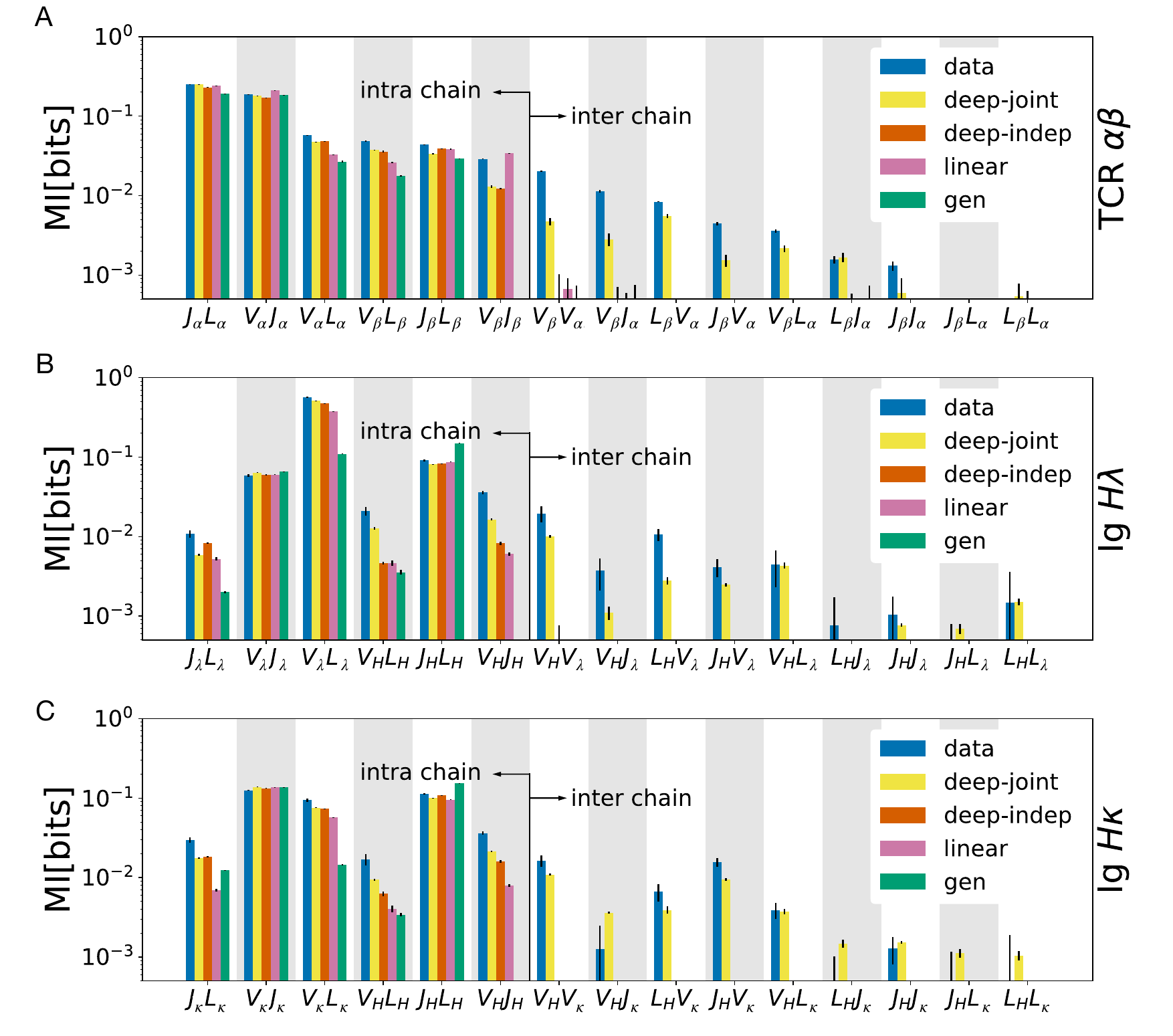}
\caption{
{\bf Inference of selection on intra- and inter-chain receptor features.} Mutual information between pairs of major intra- and inter- chain features ($V$ and $J$ gene choice and $L=$CDR3 length for each chain) for {\bf(A)} TCR $\alpha\beta$, {\bf(B)} Ig H$\lambda$, and {\bf(C)} Ig H$\kappa$ paired chains are shown. Mutual information is estimated directly from data (blue), and from receptors generated based on inferred models: generative baseline (green), \textit{linear} SONIA (pink), \textit{deep-indep} (red), and \textit{deep-joint} (yellow).
For both TCRs and BCRs, only the \textit{deep-joint} model (yellow), which correlates the features of both chains through a deep neural network, is able to recover inter-chain correlations. Mutual informations are corrected for finite-size bias and error bars are obtained by subsampling (\seemethods). {The diversity of the paired-chain B- and T-cell repertoires and the contributions of different features to this diversity are reported in Table S1.}
\label{fig:3}
}
\end{center}
\end{figure*}

Both TCRs and BCRs have intra- and inter chain correlations of sequence features, with a stronger empirical mutual dependencies present within chains  (Fig.~\ref{fig:3}). 

To account for these dependencies between chains, we generalize the selection model of eq.~\ref{eq:model} to pairs, $\x=(\x^a,\x^b)$, where $(a,b)=(\alpha,\beta)$ in TCRs or $({\rm H},\kappa)$ or $({\rm H},\lambda)$ in BCRs:
\begin{equation}
 P_{\post}(\x)= \frac{1}{Z^\theta} P_{\gen}^a(\x^a)P_{\gen}^b(\x^b) Q(\x),
\end{equation}
where we have dropped the dependence on parameters $\theta$ for ease of notation.

Analogously to single chains, we first define a \textit{linear} selection model specified by $Q(\x)=\exp(\sum_{f}\theta_fx_f)$, where the sum now runs over features of both chains $a$ and $b$. Because of its multiplicative form, selection can then be decomposed as the product of selection factors for each chain: $Q(\x)=Q^a(\x^a)Q^b(\x^b)$, where $Q^a$ and $Q^b$ are linear models. We also define a deep independent model (\textit{deep-indep}), which has the multiplicative form $Q(\x)=Q^a(\x^a)Q^b(\x^b)$, but where $Q^a$ and $Q^b$ are each described by deep neural networks that can account for complex correlations between features of the same chain, similar to the single-chain case (Fig.~S3). 
The resulting post-selection distributions for  both the  {linear} and the {deep-indep}  model factorize, $P_{\rm post}(\x)=P_{\rm post}^a(\x^a)P_{\rm post}^b(\x^b)$, making the two chains independent. Thus, by construction neither the {linear} nor the {deep-indep} model can account for correlations between chains.
Finally, we define a full soNNia model (\textit{deep-joint}) where $Q(\x)$ is a neural network combining and correlating the features of both chains (Fig.~S4). 

We trained these three classes of models on each of the TCR $\alpha-\beta$, and BCR H-$\kappa$ and H-$\lambda$ paired repertoire data described earlier. We then used these models to generate synthetic data with a depth similar to the real data, and calculated mutual informations between pairs of features (Fig.~\ref{fig:3}).
The pre-selection generation model ($Q(\x)=1$, green bars) explains  part but not all of the intra-chain feature dependencies, for both T- and B-cells, while the {linear} (purple), {deep-indep} (red), and { deep-joint} (yellow) models explain them very well. By construction, the generation, linear, and deep-indep models do not allow for inter-chain correlations. Only the { deep-joint} model (yellow) is able to recover part of the inter-chain dependencies observed in the data. It even overestimates some correlations in BCRs, specifically between the CDR3 length distributions of the two chains, and between the heavy-chain $J$ and the light-chain CDR3 length. Thus, the deep structure of soNNia recapitulates both intra-chain and inter-chain dependencies of feature forming immune receptors.

{The inferred selection on correlated inter-chain receptor features is consistent with previous analyses in TCRs~\cite{Grigaityte213462,Dupic2019,Carter2019,Tanno2020,Shcherbinin2020}  and is likely due to the synergy of the two chains interacting with self-antigens presented during thymic development for TCRs and pre-peripheral selection (including central tolerance) for BCRs, or later when recognizing antigens in the periphery. Notably, the largest inter-chain dependencies and synergistic selection are associated with the V-gene usages of the two chains (Fig.~\ref{fig:3}), which encode a significant portion of antigen-engaging regions in both TCRs and BCRs.  
 }

{Our results show that the process of selection in BCRs is restrictive, in agreement with previous findings~\cite{Nemazee2017}, significantly increasing inter-chain feature correlations. Notably, the increase in correlations (difference between green and other bars) due to selection is larger in naive B-cells than in unsorted (memory and naive) T-cells. However, the selection strengths inferred by our models should not be directly compared to estimates of the percentage of cells passing pre-peripheral selection, $\sim 10\%$ for B cells versus $3-5\%$ for T cells~\cite{Yates2014}. Our models identify features under selection without making reference to the number of cells carrying these features. Since the T-cell pool in our analysis is a mixture of naive and memory cells, we can expect stronger selection pressures in the T-cell data than in the purely naive T cells. However, previous work analysing naive and memory TCRs separately using linear selection models did not report substantial differences between the two subsets~\cite{Elhanati2014}. 

Lastly, to quantify  the diversity of immune receptor repertoires, we compared the  entropy of unpaired and paired chain repertoires in Table~S1 (\seemethods). These entropy measures suggest a repertoire size (i.e., a typical number of amino acid sequences) of about  $10^9$ receptors for TCR$\beta$ (consistent with ref.~\cite{Sethna2020}), $10^7$ receptors for TCR$\alpha$, $10^{13}$ receptors for BCR heavy chain, and $10^{4}$  receptors for BCR light chain sequences. The paired chain entropy  measures suggest  repertoire sizes of $10^{16}$ for TCR$\alpha\beta$ and $10^{17}$ BCR IgH$\lambda$ and IgH$\kappa$ receptors, which are compatible with the small correlations observed between heavy and light chains in Fig.~3, and previously reported in refs.~\cite{Grigaityte213462,Dupic2019,Carter2019,Tanno2020,Shcherbinin2020}.
}

\label{sec_paired}

\subsection*{Cell type and tissue-specific selection on T-cells} 

\begin{figure*}[t!]
\begin{center}\includegraphics[width=0.9\linewidth]{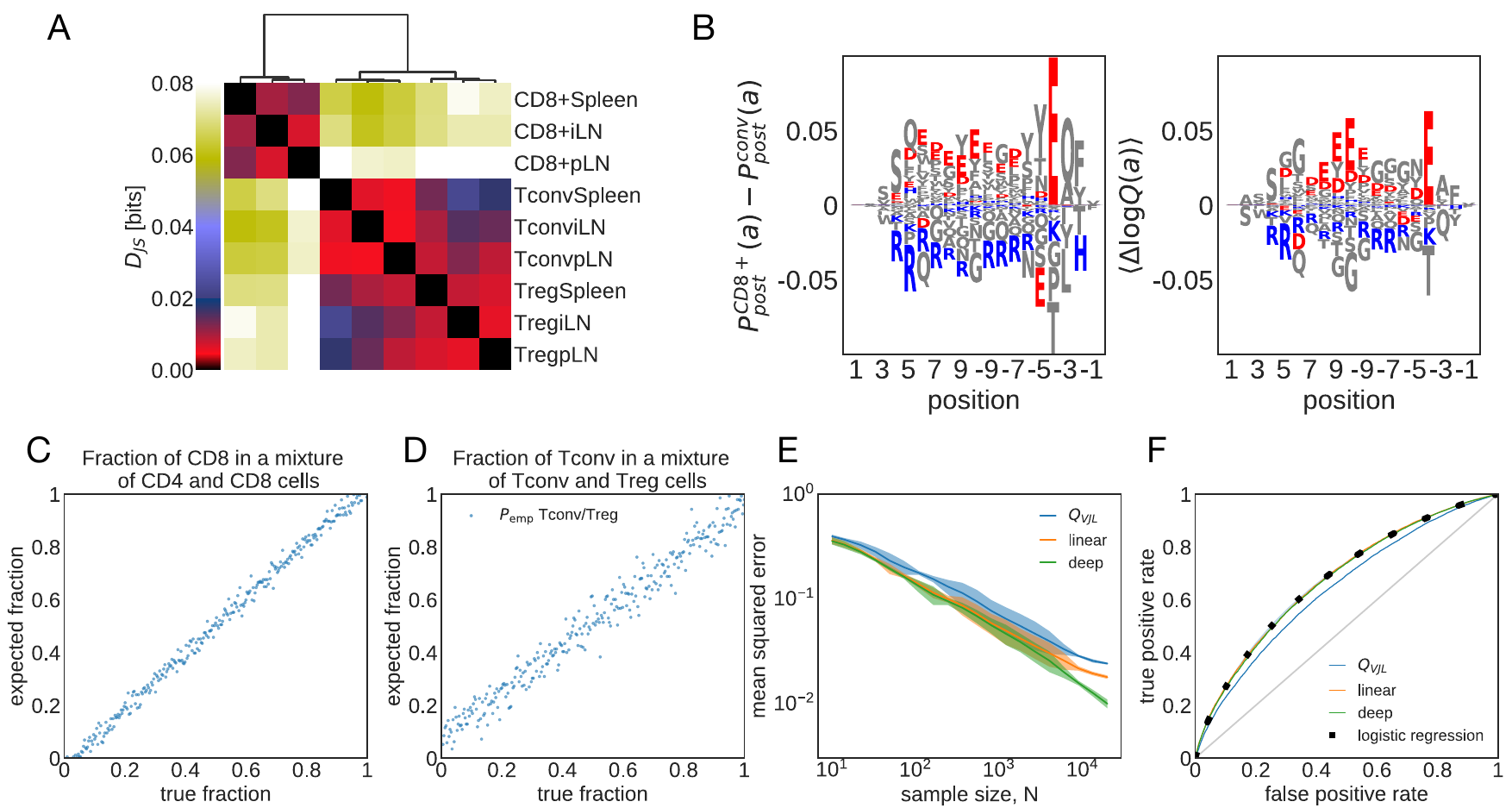}
\caption{{\bf Cell type and tissue-specific selection on TCRs.} 
{\bf(A)} Jensen-Shannon divergences ($D_{JS}$, see eq.~\ref{eq:djs}) computed from models trained on different sub-repertoires are shown. 
{{\bf (B)} 
Difference in the marginal probability  for amino acid composition along the CDR3, $P^{\text{CD8}}_{\post}(a)-P^{\text{CD4}}_{\post}(a)$, between CD8$^+$ and CD4$^+$ Tconv (left) and the mean difference in the corresponding log-selection factors for amino acid usage  $\Delta \log Q = \log Q^{\text{CD8}} - \log Q^{\text{CD4}}$ (right)  are shown (the mean is taken over the distribution $(P^{\text{CD8}}_{\post}+P^{\text{CD4}}_{\post})/2$).
The negatively charged amino acids (Aspartate, D, and Glutamate, E) and the positively charged amino acids (Lysine, K, and Arginine, R) are indicated in red and blue, respectively. Other amino acids are shown in gray.}
{\bf(C)} Maximum-likelihood inference of the fraction of CD8$^+$ TCRs in mixed repertoires of conventional CD4$^+$ T cells (Tconvs) and CD8$^+$ cells from spleen (Eq~\ref{eq:4}) is shown. Each repertoire comprises $5\times 10^3$ unique TCRs. {\bf(D)} Same as (C) but for a mixture of Tconv and Treg TCRs. {\bf(E)} Mean squared error of the inferred sample fraction from (C) as a function of sample size $N$, averaged over all fractions, using models of increasing complexity: ``$Q_{VJL}$" is a linear model with only features for CDR3 length and VJ usage, ``linear" is linear SONIA model, ``deep" is the full soNNia model (Fig.~\ref{emersonfig}C).
  {\bf(F)} Receiving-Operating Curve (ROC) for classifying individual sequences coming from CD8$^+$ cells or from CD4$^+$ Tconvs from spleen, using the log-likelihood ratios.
  Curves are generated by varying the threshold in eq.~\ref{eq:classifier}. The accuracy of the classifier is compared to a traditional logistic classifier inferred on the same set of features as our selection models. The training set for the logistic classifier has $N=3 \times 10^5$ Tconv CD4$^+$, and $N= 8.7 \times 10^4$ CD8$^+$ TCRs, and the test set has $N= 2 \times 10^4$ CD4$^+$, and $N= 2 \times 10^4$ CD8$^+$ TCR sequences.}
\label{fig:compare}
\end{center}
\end{figure*}

During maturation in the  thymus, T-cells differentiate into two major cell-types: cytotoxic (CD8$^+$) and helper (CD4$^+$) T-cells. CD8$^+$ cells bind peptides presented on major histocompatibility complex (MHC) class I molecules that are expressed by all cells, whereas CD4$^+$ cells bind peptides presented on MHC-class II molecules, which are only expressed on specialized antigen presenting cells. Differences in sequence features of CD8$^+$ and CD4$^+$ T-cells should reflect the distinct recognition targets of these receptors. Although these differences have already been investigated in refs.~\cite{Seay2016,Carter2019}, we still lack an understanding as how selection contributes to the differences between CD8$^+$ and CD4$^+$ TCRs. In addition to functional differentiation at the cell-type level, T-cells also migrate and reside in different tissues, where they encounter different environments  and are prone  to infections by different pathogens. As a result, we expect to detect tissue-specific TCR preferences that reflect tissue-specific T-cell signatures. 

To characterize differential sequence features of TCRs between cell types  in different tissues,  we  pool unique TCRs from  9 individuals (from ref.~\cite{Seay2016}) sorted into three cell-types (CD4$^+$ conventional T cells (Tconv), CD4$^+$ regulatory T cells (Treg) and  CD8$^+$ T cells), and harvested from 3  tissues (pancreatic draining lymph nodes (pLN), ``irrelevant" non-pancreatic draining lymph nodes (iLN), and spleen).

Training a nonlinear soNNia model (see Fig.~\ref{emersonfig}~C) for each subset leads to overfitting issues due to limited data. {To solve this problem, we train the model in two steps. First, we use the unfractionated data from ref.~\cite{Dean2015} to construct a shared baseline for all repertoire subsets. We then learn independent linear SONIA models for each repertoire subset so that the inferred $Q$ factors only reflect selection relative to the baseline. We approach this problem in two ways: (i)  We infer a SONIA model atop an empirical  baseline set $\G$ constructed from the  unfractionated repertoire, and (ii) we use the technique of transfer learning,  which consists of learning a shared nonlinear soNNia model for the unfractionated repertoire and then add an additional linear layer (similar to standard SONIA) for each sub-repertoire  (see SI and Fig.~S5). The  sub-repertoire selection factors $Q$ inferred by these two approaches are very similar (Fig.~S5), but the former method is simpler and we use it for our main analysis in Fig.~4. For comparison, we also used the generation model $P_{\rm gen}$ (trained earlier for Fig.~2) as a baseline, in which case the selection factors include selection effects that are shared among the sub-repertoires. Distributions of selection factors obtained by both approaches are shown in Fig.~S6.
}

{To quantify differential selection on sub-repertoires we use Jensen-Shannon divergence $D_{\rm JS}$ between the distributions of receptors $P_\post^{r}$ and $P_\post^{r'}$ for  pairs of  sub-repertoires ($r,\, r'$)    (Methods).} Clustering of cell types based on Jensen-Shannon divergence shows strong differential selection preferences between the CD4$^+$ and CD8$^+$ receptors, with an average $D_{\rm JS}\simeq 0.08\pm 0.01$ bits across respective tissues and sub-repertoires  (Fig.~\ref{fig:compare}A; see also Fig.~S7A for similar results where $P_{\rm gen}$ is used as baseline). We identify differential selection between Tconv and T{reg} receptors within CD4$^+$ cells with $D_{JS}\simeq 0.015\pm0.004$. We also detect moderate tissue specificity for CD8$^+$ and Treg receptors, but no such signal can be detected for CD4$^+$ Tconv cells across different tissues.

Examining the linear selection factors of the SONIA model trained atop $P_{\rm gen}$ as a baseline reveals the VJ (Fig.~S8) and amino-acid usage features (Fig.~\ref{fig:compare}C) that are differentially selected in the Tconv CD4$^+$ and CD8$^+$ subsets (in spleen). 
Linear selection models are organised according to a hierarchy from the least to the most constrained model. As one adds selection factors for each feature, the Kullback-Leibler divergence between the repertoire and the baseline increases (\seemethods). Decomposing in this way the divergence between CD4$^+$ Tconv and CD8$^+$ repertoires, we find that contributions to the total divergence are evenly split between amino-acid features and VJ gene usage, with only a minor contribution from CDR3 length (Fig.~S9). {It should be noted that the baseline models $P_\gen$  for these sub-repertoires, inferred from their unproductive receptors, are similar (Fig.~S10) and do not contribute to these differential preferences.}

{One key difference between CD4$^+$ and CD8$^+$  TCRs amino acid composition is their CDR3 charge preferences. We observe an over-representation of positively charged (Lysine, K, and Arginine, R) and suppression of negatively charged (Aspartate, D, and Glutamate, E) amino acids in CD4$^+$ TCRs compared to CD8$^+$ TCRs (Fig.~\ref{fig:compare}B), consistent with previous observations~\cite{Li2016a}. These charge preferences arise due to differential selection on the two subtypes (Fig.~\ref{fig:compare}B), indicating  broad differences between amino acid features of peptide-MHC-I and  peptide-MHC-II complexes, which respectively interact with  CD8$^+$ and  CD4$^+$ TCRs.  For example, a statistical survey of peptides presented by different MHC classes show that  MHC-I molecules tend to present more positively charged  peptides compared to  MHC-II molecules---a bias that is complementary to the charge preferences of the respective TCR subtypes~\cite{Rapin2010}.}

\subsection*{Decomposing unsorted repertoires using selection models} Knowing $P^{r}_\post$ models specific to sub-repertoires enables us to infer the fraction of each class $r$ in unsorted data. Estimating the relative fraction of CD4$^+$ and CD8$^+$ sub-types in a repertoire can be informative for clinical purposes, e.g. as a probe for Tumor Infiltrating Lymphocytes (TIL), where over-abundance of CD8$^+$ cells in the sample has been associated with positive prognosis in ovarian cancer~\cite{Sato18538}. Given a repertoire composed of the mixture of two sub-repertoires $r$ and $r'$ in unknown proportions, we maximize the log-likelihood function $L(f)$ based on our selection models to find the fraction $f$ of a sub-repertoire $r$ within the mixture:
\begin{align}\label{eq:4}
L(f) &=\langle\log(f P^r_\post (\sigma) +(1-f)P^{r'}_\post (\sigma))\rangle_D \\
	  &= \langle \log( f \mathcal{Q}^r(\sigma) +(1-f)\mathcal{Q}^{r'}(\sigma))\rangle_D + {\rm const},\nonumber
\end{align}
where $\langle \cdot\rangle_{D}$ is the empirical mean over sequences in the mixture.

Previous work has used differential V- and J-gene usage, and CDR3 length to characterize the relative fraction of CD4$^+$ and CD8$^+$ cells in an unfractionated repertoire~\cite{EMERSON201314}.  The log-likelihood function in eq.~\ref{eq:4} provides a principled approach for inferring cell-type composition using selection factors that capture the differential receptor features of each sub-repertoire, including but not limited to their V- and J- usage and CDR3 length and amino acid preferences.

To test the accuracy of our method, we formed a synthetic mixture of previously sorted CD4$^+$ (Tconv from spleen~\cite{Seay2016}) and CD8$^+$ (from spleen~\cite{Seay2016}) receptors with different proportions, and show that our selection-based inference can accurately recover the relative fraction of CD8$^+$ in the mix (Fig \ref{fig:compare}C). Our method can also infer the proportion of Treg cells in a mixture of Tconv and Treg CD4$^+$ cells from spleen (Fig.~\ref{fig:compare}D), which is a much harder task since these subsets are very similar (Fig.~\ref{fig:compare}A). The accuracy of the inference depends on the size of the unfractionated data, with a mean expected error  that falls below 1\% for datasets with  size $10^4$ or larger for the CD8$^+$/CD4$^+$ mixture (red and orange lines in Fig.~\ref{fig:compare}E). 

Our method uses a theoretically grounded maximum likelihood approach, which includes all the features captured by the soNNia model. Nonetheless, a simple linear selection model with only V- and J- gene usage and CDR3 length information (blue line in Fig.~\ref{fig:compare}E), analogous to the method used in ref.~\cite{EMERSON201314}, reliably infers the composition of the mixture repertoire. Additional information about amino acid usage in the linear SONIA model results in moderate but significant improvement  (orange line). The accuracy of the inference is insensitive  to the choice of the baseline model for receptor repertoires: using the empirical baseline from ref.~\cite{Dean2015} (Fig.~\ref{fig:compare}E) or  $P_{\rm gen}$ (Fig.~S7D) does not substantially change the results.

The method can be extended to the decomposition of 3 or more sub-repertoires. To illustrate this, we inferred the fractions of Tconv, Treg, and CD8$^+$ cells in synthetic unfractionated repertoires from spleen, showing an accuracy of $3 \pm 1 \%$ in reconstructing all three fractions (Fig.~S11) in a mixture of size $5\times10^3$.

\begin{figure*}[t!]
\begin{center}
\includegraphics[width=0.8\linewidth]{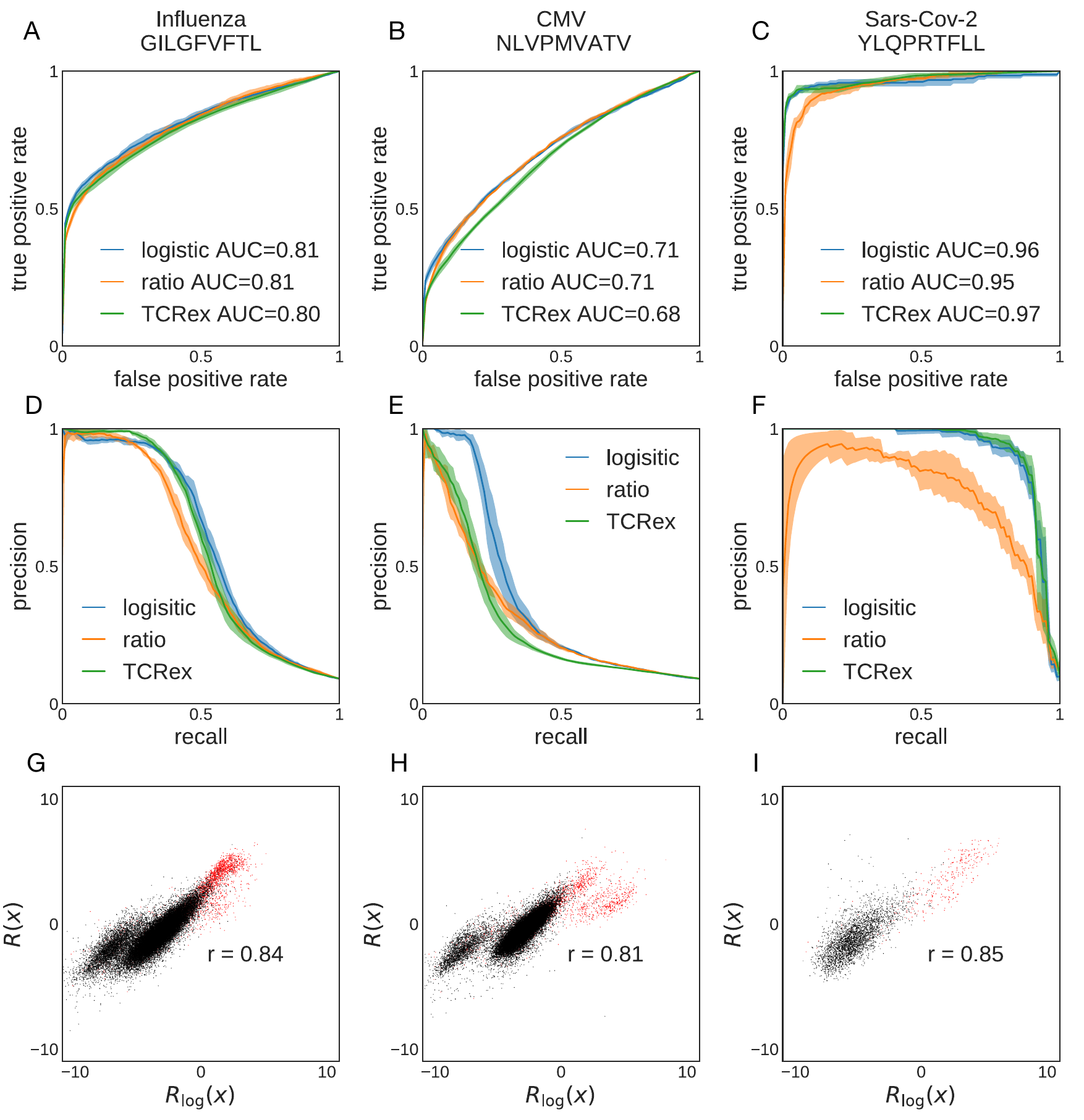}
\caption{
 {\bf Selection-based prediction of epitope specificity for TCR.}
 TCRs are classified based on their reactivity to three pathogenic epitopes (columns), using three classification methods: TCRex, log-likelihood ratio (Eq.~\ref{eq:classifier}), and linear logistic regression (Eq.~\ref{eq:loglin}). {\bf(A-C)} ROC curves, and {\bf(D-F)} precision-recall curves for {\bf (A,D)} influenza epitope GILGFVFTL ($N=3107$ TCR), {\bf (B,E)} CMV epitope NLVPMVATV ($N=4812$), and {\bf (C,F)} SARS-CoV-2 epitope YLQPRTFLL ($N=315$) are shown. {\bf (G-I)} Comparison between log-likelihood scores $R(x)$  and logistic regression scores  $R_{\rm log}(x)$, for the three epitopes. Red points are TCRs that bind the specific epitope (positive set), black points are TCRs from bulk sequencing (negative set). $r$ is Pearson's correlation.
  For all panels we used pooled data from Ref.~\cite{Dean2015} as the negative set. We used 10 times more negative data than positive data for training. Performance was quantified using 5-fold cross-validation.
\label{fig:classify}
}
\end{center}
\end{figure*}

\subsection*{Computational sorting of CD4$^+$ and CD8$^+$ TCR}
Selection models are powerful in characterizing the broad statistical differences between distinct functional subsets of immune receptors, including the CD4$^+$ and CD8$^+$ TCRs (Fig.~\ref{fig:compare}A). A more difficult task, which we call computational sorting, is to classify {\em individual} receptors into functional classes based on their sequence features. {In other words, how accurately can one classify a given receptor as a member of a functional subset (e.g. CD4$^+$ or CD8$^+$ TCRs)?}

 We use selection models inferred for distinct sub-repertoires  $r$ and $r'$ to estimate a log-likelihood ratio $R(\x)$ for a given receptor 
  $\x$  to belong to  either of the sub-repertoires,
\begin{equation}
R(\x)=\log\frac{P_{\rm post}^{r}(\x)}{P_{\rm post}^{r'}(\x)}= \log\frac{\mathcal{Q}^r(\x)}{\mathcal{Q}^{r'}(\x)}.
\label{eq:classifier}
\end{equation}
A larger log-likelihood ratio $R(\x)$ indicates that the receptor is more likely to be associated with  the sub-repertoire $r$ than $r'$. We set a threshold $R_c$, to assign a receptor to $r$ if $R(\x)\geq R_c$ and to $r'$ otherwise. The sensitivity and specificity  of this classification depends on the threshold value. We evaluate the accuracy of our log-likelihood classifier between sets of CD8$^+$ and Tconv CD4$^+$ receptors harvested from spleen~\cite{Seay2016}. The Receiver Operating Characteristic (ROC) curve in Fig.~\ref{fig:compare}F shows that our selection-based method can classify receptors as  CD8$^+$ or CD4$^+$ cells, with an area under the curve AUC = 0.68. Performance does not depend on the  choice of the baseline model ($P_{\rm emp}$ in Fig.~\ref{fig:compare}F and $P_{\rm gen}$ in Fig.~S7E). Applying this classification method to all the possible pairs of sub-repertoires in Fig.~\ref{fig:compare}A, we find that CD4$^+$ vs CD8$^+$ discrimination generally achieves AUC$\approx 0.7$, while discriminating sub-repertoires within the CD4$^+$ or CD8$^+$ classes yields much poorer performance (Fig.~S12).

For comparison, we also used a common approach for categorical classification and optimized a linear logistic classifier that takes receptor features (similar to the selection model) as input, and classifies receptors into CD8$^+$ or CD4$^+$ cells. The model predicts the probability that sequence $\x$ belongs to sub-repertoire $r$ (rather than $r'$) as $\hat{y}(\mathbf{x})=\zeta(R_{\rm log}(\x))$, with $R_{\rm log}(\x)=\sum_{f} w_f x_f+b$ and $\zeta(x)=e^x/(1+e^x)$. We learn the model parameters $w_f$ and $b$ by maximizing the log-likelihood of the training set:
\begin{equation}
\mathcal{L}_c(\mathbf{w},b)=\sum_{i=1}^N\Big[y_i\log\hat{y}(\mathbf{x})
+(1-y_i)\log(1-\hat{y}(\mathbf{x}_i))\Big]
\label{eq:loglin}
\end{equation}
where $y_i$ labels each TCR by their sub-repertoire, e.g. $y_i=1$ for CD8$^+$, and $y_i=0$ for CD4$^+$. 
Note that when selection models are linear, the log-likelihood ratio (eq.~\ref{eq:classifier}) also reduces to a linear form---the only difference being how the linear coefficients are learned. This optimized logistic classifier (eq.~\ref{eq:loglin}) performs equally well compared to the selection-based classifier, with the same AUC=0.68 (points in Fig.~\ref{fig:compare}F). These AUCs  are comparable to those found in ref.~\cite{Carter2019}, which has addressed the same issue using black-box machine learning approaches.

It should be emphasized that despite comparable performances, our fully linear selection-based  method provides a biologically interpretable basis for subtype classification, in contrast to black box approaches~\cite{Carter2019}. {For example, the relative importance of different sequence features (i.e., CDR3 length, V / J gene identity and amino acid composition) for CD4$^+$ vs. CD8$^+$ classification  are shown in Fig.~S9.}

\subsection*{Classification of TCRs targeting distinct antigenic epitopes} 
Recognition of a pathogenic epitope by a TCR is mediated through molecular interactions between the two proteins. The strength of this interaction depends on the complementarity of a TCR against an antigen presented by a MHC molecule on the T-cell surface. 
Recent growth of data on paired TCRs and their target epitopes~\cite{Shugay2018,Bagaev2019} has led to the development of machine learning methods for TCR-epitope mapping~\cite{Glanville2017,Shugay2018,Jokinen2019a,Gielis2019,Dash2017}. A TCR-epitope map is a classification problem that determines whether  a TCR binds to a specific epitope. We use our selection-based classifier (eq.~\ref{eq:classifier}) to address this problem. We determine the target ensemble $P_{\rm post}^r$ from the training set of TCRs associated with a given epitope (positive data), and the alternative ensemble $P_{\post}^{r'}$ from a set of generic unfractionated TCRs (negative data). For comparison, we also perform the classification task using the linear logistic regression approach (eq.~\ref{eq:loglin}), and the state of the art TCRex algorithm~\cite{Gielis2019}, which uses a random forest model for classification.

We  performed classification for the following CD8$^+$-specific epitopes, presented on HLA-A*02 molecules: (i) the influenza GILGFVFTL epitope (with $N=3107$ associated TCRs), (ii) the Cytomegalovirus (CMV) NLVPMVATV epitope ($N=4812$), and (iii) the SARS-CoV-2 YLQPRTFLL epitope ($N=315$). The first two epitopes have the most abundant associated TCR sets in VDJdb~\cite{Shugay2018,Bagaev2019}, and the latter is relevant for the ongoing COVD-19 pandemic. For consistency with TCRex~\cite{Gielis2019}, we used the pooled data from ref.~\cite{Dean2015} as the negative set, and used 10 times more negative data than positive data for training. To quantify performance of each classifier, we performed a 5-fold cross validation procedure. Due to the scarcity of data, we limit our selection inference to the  linear SONIA model (see Fig.~\ref{emersonfig}C). The ROC curves show comparable performances for the three classification methods  on the three epitope-specific TCR sets (Fig.~\ref{fig:classify}A-C).

The TCR-epitope mapping is a highly unbalanced classification problem, where reactive receptors against a specific epitope comprise a very small fraction of the repertoire (less than $10^{-5}$~\cite{Yates2014}). Precision-recall curves are best suited to evaluate the performance of classification for imbalanced problems. In this case, a classifier should show a large  precision (fraction of true predicted positives among all predicted positives) for a broad range of recall or sensitivity (fraction of true predicted positives among positives = true positives + false negatives). The precision-recall curves in Fig.~\ref{fig:classify}D-F show that TCRex and the logistic classifier can equally well classify the data, and moderately outperform the selection-based classifier. While both the logistic classifier and TCRex are optimized for classification tasks, the selection-based classifier is a {\it generative} model trained to infer the receptor distribution of interest (positive set) and identify its distinguishing features from the baseline (negative set). As a result, selection-based classification underperforms in the low-data regime, for which fitting a reliable distribution is difficult  (e.g. for the SARS-CoV-2 epitope model, with only $N=315$ positive examples). By contrast, the logistic classifier finds a hyperplane that best separates the two sets, and therefore, is better suited for classification tasks, and may be trained on smaller datasets. Nonetheless, we see a strong correlation between the selection-based log-likelihood ratio $R(x)$ (eq.~\ref{eq:classifier}) and the estimator of the logistic classifier $\hat y$ (eq.~\ref{eq:loglin}), shown for positive set (red points) and the negative set (black points) in Fig.~\ref{fig:classify}G-I for the three epitopes. This result indicates that the separation hyperplane identified by the logistic classifier aligns well along the effective coordinates of selection that represent sequence features relevant for function in each epitope class.

\section*{Discussion} 
Previous work has developed linear selection models to characterize the distribution of productive T cell receptors~\cite{Sethna2020}. Here, we generalized on these methods by using deep neural networks implemented in the soNNia algorithm to account for nonlinearities in feature space, and have improved the statistical characterization of TCR repertoires in a large cohort of individuals~\cite{Emerson2017}.

Using this method, we modelled the selective pressure on paired chains of T- and B- cell receptors, and found that the observed cross-chain correlations, even if limited, could be partially reproduced with our model (Fig.~3). These observed inter-chain correlations are likely due to the synergy of the two chains interacting with self and non-self antigens, which determine the selection pressure that shape the functional TCR and BCR repertoires.

We  systematically compared T cell subsets and showed that our method identifies differential selection   on CD8$^+$ T-cells, CD4$^+$ conventional T-cells, and CD4$^+$ regulatory T-cells. TCRs belonging to families with more closely related developmental paths (i.e., CD4$^+$ regulatory or conventional cells) have more similar selection features, which differentiate them from cells that diverged earlier (CD8$^+$). Cells with similar functions in different tissues are in general similar, with the exception of spleen CD8$^+$ that stands out from lymph node CD8$^+$. {These differences capture broad  differential preferences of CD8$^+$ and CD4$^+$ TCRs, which can arise from their distinct structural features complementary to their different targets, i.e.,  peptide-HLAI and  peptide-HLAII complexes.  A next step would be to uncover more fine-grained differential features, associated with the distinct pathogenic history or HLA composition of different individuals.}

One application of the soNNia method is to utilize our selection models to infer ratios of cell subsets in unsorted mixtures, following the proposal of Emerson et al.~\cite{EMERSON201314}. Consistently with previous results, we find that the estimated ratio of CD4$^+$/CD8$^+$ cells in unsorted mixtures achieves precision of the order of $1\%$ with as few as $10^4$ unique receptors. Emerson et al. validated their computational sorting based on sequence identity on data from in-vitro assays and flow cytometry, which gives us confidence that our results would also pass an experimental validation procedure.

As a harder task, we were also able to decompose the fraction of regulatory versus conventional CD4$^+$ T-cells, showing that receptor composition encodes not just signatures of shared developmental history--- receptors of these two CD4$^+$ subtypes are still much more similar to each other than to CD8$^+$ receptors--- but also function: Tregs down-regulate effector T-cells and curb an immune response creating tolerance to self-antigens and preventing autoimmune diseases~\cite{Wing2010}, whereas Tconvs assist other lymphocytes including activation of differentiation of B-cells. Since our analysis is performed on fully differentiated peripheral cells, we cannot say at what point in their development these CD4$^+$ T-cells are differentially selected. Data from regulatory and conventional T-cells at different stages of thymic development could identify how their receptor composition is shaped over time.

During thymic selection cells first rearrange a $\beta$ receptor and then an $\alpha$ receptor is added concurrently with positive selection. Negative selection follows positive selection and overlaps with CD4/CD8 differentiation. We found that the Jensen-Shannon divergence between CD8$^+$ and CD4$^+$ cells to be very small (0.1 bit) compared to the divergence between functional and generated repertoires (ranging from 0.8 to 0.9 bits). This result suggests that the selection factors captured by our model mainly act during positive selection, which is partly shared between CD4$^+$ and CD8$^+$ cells, rather than during cell type differentiation and negative selection, which is distinct for each type. Additionally to showing statistical differences in sub-repertoires, we classified cells into CD4$^+$ and CD8$^+$ subclasses with likelihood ratios of selection models and recovered similar results achieved using pure machine learning  approaches~\cite{Carter2019}, but in a fully linear and interpretable setting. 

In recent years multiple machine learning methods have been proposed in order to predict antigen specificity of TCRs: TCRex~\cite{Gielis2019,DeNeuter2018}, DeepTCR~\cite{Sidhom464107}, netTCR~\cite{Jurtz433706}, ERGO~\cite{Springer2020}, TCRGP~\cite{Jokinen542332} and TcellMatch~\cite{Fischer2020}. All these methods have explored the question in slightly different ways, and made comparisons with each other. However, with the sole exception of TcellMatch~\cite{Fischer2020}, none of the above methods compared their performance to a simple linear classifier. TcellMatch~\cite{Fischer2020} does not explicitly compare to other existing methods, but implicitly compares various neural network architectures. We thus directly compared a representative of the above group of machine learning models, TCRex, to a linear logistic classifier, and to the log-likelihood ratio obtained by training two SONIA models on the same set of features. We found that the three models performed similarly (Fig.~\ref{fig:classify}), consistent with the view that amino acids from the CDR3 loop interact with the antigenic peptide in an additive way. This result complements similar results in Ref.~\cite{Fischer2020}, where a linear classifier gave comparable results to deep neural network architectures. 

The linear classifier based on likelihood ratios achieves state-of-the art performance both in discriminating CD4$^+$ from CD8$^+$ cells (Fig.~\ref{fig:compare}F), and in predicting epitope specificity (Fig.~\ref{fig:classify}). But unlike other classifiers, its engine can be used to generate positive and negative samples. Thus characterizing the distributions of positive and negative examples is more data demanding than mere classification. For this reason pure classifiers are generally expected to perform better, but lack the ability to sample new data. Our analysis complements the collection of proposed classifiers by adding a generative alternative that is grounded on the biophysical process of T-cell generation and selection. This model is simple and interpretable, and performs well with large amounts of data.

The epitope discrimination task discussed here and in previous work focuses on predicting TCR specificity to one specific epitope. A long-term goal would be to predict the affinity of any TCR-epitope pair. However, currently available databases \cite{Shugay2018,Bagaev2019} do not contain sufficiently diverse epitopes to train models that would generalize to unseen epitopes \cite{Fischer2020}. A further complication is that multiple TCR specificity motifs may co-exist even for a single epitope \cite{Dash2017,Minervina2020}, which cannot be captured by linear models \cite{Bravi2020}. Progress will be made possible by a combination of high-throughput experiments assaying many TCR-epitope pairs \cite{Klinger2015}, and machine learning based techniques such as soNNia.

In summary, we show that nonlinear features captured by soNNia capture more information about the initial and peripheral selection process than linear models. However, deep neural network methods such as soNNia suffer from the drawback of being data hungry, and show their limitations in practical applications where data are scarce. {Nonetheless, with the rapid growth of functionally annotated datasets, we expect soNNia to be  more readily used for  inference of nonlinear selection  on immune receptor sequences. Such nonlinearity is expected as it would reflect the ubiquitous epistatic interactions between residues of a receptor protein that encode for a specific function.} In a more general context, soNNia is a way to integrate more basic but interpretable knowledge-based models and more flexible but less interpretable deep-learning approaches within the same framework. 

{
 \section*{Methods}
 {
 {\small 
\noindent {\bf Data description.}  In this work we used different datasets to  evaluate selection on T- and B-cell receptor features. 
\begin{enumerate}
\item To quantify the accuracy of the soNNia model (Fig.~\ref{emersonfigB}), we used the TCR$\beta$ repertoires from a large cohort of  743 individuals from ref.~\cite{Emerson2017}. We pool the {\it unique} nucleotide sequences of receptors from all individuals and construct a universal donor totalling $9\times 10^7$ sequences. We randomly split the pooled dataset into a training and a test set of equal sizes. We then subsampled the training set to $10^7$ to reduce the computational cost of inference .
\item To characterize selection on paired chain receptors (Fig.~\ref{fig:3}), we analyzed TCR $\alpha\beta$ pairs of unfractionated repertoires from ref.~\cite{Tanno2020} (totalling $5\times 10^5$ receptors), and BCR of naive cells from ref.~\cite{DeKoskyE2636} totalling $22\times10^3$ and $28\times10^3$ receptors for the H$\lambda$ and H$\kappa$ repertoires, respectively.
\item To characterize differential sequence features of TCRs between cell types in different tissues (Fig.~\ref{fig:compare}), we pooled unique TCRs from 9 healthy individuals from ref.~\cite{Seay2016}, sorted into CD4$^+$ conventional T cells (Tconv), CD4$^+$ regulatory T cells (Treg) and CD8$^+$ T cells, harvested from 3 tissues: pancreatic draining lymph nodes (pLN) ($2.3\times10^5$ Tconvs, $2.9\times10^5$ Tregs, $2.5\times10^5$ CD8s), ``irrelevant" non-pancreatic draining lymph nodes (iLN) ($2.0\times10^5$ Tconvs, $9.0\times10^4$ Tregs, $1.0\times10^5$ CD8s), and spleen  ($3.2\times10^5$ Tconvs, $1.1\times10^5$ Tregs, $1.1\times10^5$ CD8s). We used the unfractionated data from ref.~\cite{Dean2015}, comprising of $2.2\times10^6$ receptor to construct a based line model for this analysis. \\
\end{enumerate}

\noindent {\bf Quantifying accuracy of selection models.}  To assess the  performance of our selection models, we compare their inferred probabilities $P_\post (\x)$ with the observed frequencies of the receptor sequences $P_{\rm data}(\x)$ in the test set. Prediction accuracy can be quantified through the Pearson correlation between the two log-frequencies or the Kullback-Leibler  divergence between the data and the distribution predicted by the selection model $P_{\rm post}$, \begin{equation}
\mathcal{D}_{\rm KL}(P_{\rm data}|P_{\rm post})=\left\langle \log_2\frac{P_{\rm data}}{P_{\rm post}}\right\rangle_{P_{\rm data}}.
\end{equation}
A smaller Kullback-Leibler divergence indicates a higher accuracy of the inferred model in predicting the data. In Fig.~\ref{emersonfigB} we estimate the Kullback-Leibler divergence using $10^5$ receptors in the test set with multiplicity larger than two. \\

\noindent {\bf Comparing selection on different sub-repertoires.} To characterize differences in sub-repertoires due to selection,  we evaluate the Jensen-Shannon divergence $D_{\rm JS}(r,r')$ between the distribution of pairs $(r,r')$ of sub-repertoires, $P_\post^{r}$ and $P_\post^{r'}$,
\begin{equation}
D_{\rm JS}(r,r')=\frac{1}{2}\Big\langle \log_2 \frac{2\Q^{r}}{\Q^{r} +\Q^{r'}} \Big\rangle_{r}
+ \frac{1}{2} \Big\langle \log_2 \frac{2\Q^{r'}}{\Q^{r} +\Q^{r'}} \Big\rangle_{r'}
\label{eq:djs}
\end{equation}
where $\langle\cdot\rangle_{r} $ denotes averages over $P_\post^r$ (\seemethods\ for evaluation details). This divergence is symmetric and only depends on the relative differences of selection factors between functional sub-repertoires, and not on the baseline model.
}}}

\end{document}


\section*{Supporting Information}
{\large \bf Deep generative selection models of T and B cell receptor repertoires with soNNia}\\
 Giulio Isacchini, Aleksandra M. Walczak, Thierry Mora, Armita Nourmohammad

\vspace{0.5cm}
\tableofcontents

\vspace{0.5cm}

\renewcommand{\theequation}{S\arabic{equation}}

\section{SoNNia}

SoNNia is a python software which extends the functionality of the SONIA package, using deep neural network architectures. It expands the choice of selection models to infer, by adding non-linear single-chain models and non-linear paired-chain models. 
\comment{Like other deep neural network algorithms, soNNia is powerful when trained on large datasets. While the use of appropriate regularization could reduce the risk of overfitting, it is recommended that the linear SONIA model is used for datasets with fewer than $10^5$ receptor sequences.}

The pre-processing pipeline implemented in this paper is also included in the soNNia package as a separate class.
The software is available on GitHub at \href{https://github.com/statbiophys/soNNia}{https://github.com/statbiophys/soNNia}. 

\section{Pre-processing steps}
The standard pre-processing pipeline, which is implemented in the soNNia package and is applied to all datasets, consists of the following steps:

\begin{enumerate}
\item Select species and chain type
\item Verify sequences are written as V gene, CDR3 sequence, J gene and remove sequences with unknown genes and pseudogenes
\item Filter  productive CDR3 sequences (lack of stop codons and nucleotide sequence length is a multiple of 3)
\item Filter sequences starting with a cysteine
\item Filter sequences with CDR3 amino acid length smaller than a maximum value (set to 30 in this paper)
\item Remove sequences with small read counts (optional).
\end{enumerate}

For the analysis of Fig.~2 we analysed data from~\cite{Emerson2017}. We first applied the standard pipeline. In addition we excluded TCRs with gene TRBJ2-5 which is badly annotated by the Adaptive pipeline \cite{Davidsen2018} and removed a cluster of artefact sequences, which was previously identified in  \cite{DeWitt2018c} and corresponds to the consensus sequence CFFKQKTAYEQYF.\\
  
For the analysis of Fig.~3 we analysed data from \cite{Tanno2020} and \cite{DeKoskyE2636}. Dataset from ~\cite{Tanno2020} was obtained already pre-processed directly from the authors, while pre-processed dataset from \cite{DeKoskyE2636} is part of the supplementary material of the corresponding paper. The soNNia standard pipeline is then applied to both datasets, independently for each chain, and a pair is accepted only if it passes both filtering steps. For $\alpha$ TCR datasets, sequences carrying the following rare genes were removed due to their rarity in the out-of-frame dataset: TRAJ33, TRAJ38, TRAJ24, TRAV19.\\

For the analysis of Figs.~4 and 5 we analysed data from \cite{Seay2016} and \cite{Dean2015}, to which we applied our standard pre-processing pipeline.
  
\section{Generation model}
The generation model relies on previously published models described in \cite{Murugan2012,Marcou2018,Sethna2019}. Briefly, the model is defined by the probability distributions of the various events involved in the VDJ recombination process: V, D, and J gene usage, and number of deletions and insertions at each junction. The model is learned from non-productive sequences using the IGoR software \cite{Marcou2018}. For BCR, only a few nonproductive sequences were available, and so we instead started from the default IGoR models learned elsewhere \cite{Marcou2018}, and re-inferred only the V gene usage distribution for the heavy chain, and VJ joint gene distribution for light chains, keeping all other parameters fixed.\\

Amino-acid sequence probability computation and generation is done with the OLGA software, which relies on a dynamic
programming approach. The process is applied to all $\alpha$, $\beta$, IgH and Ig$\kappa$/$\lambda$ chains. 
We focus on naive B cells and ignore somatic hypermutations. Since it was shown that individual variability in generation was only small \cite{Sethna2020}, for each locus we used a single universal model.\\

\section{Neural network architectures}
We  describe the architecture of the soNNia neural network.  
The input of our network is a vector $\mathbf{x}$ where  $x_f=1$ (otherwise 0) if sequence $x$ has feature $f$. A dense layer is a map $\mathbf{L}(\mathbf{x})=\tanh(\mathbf{W}\mathbf{x}+\mathbf{b})$ with $\mathbf{x}$ the input vector, $\mathbf{W}$ the matrix of weights, $\mathbf{b}$ the vector bias, and where the $\tanh$ function is applied to each element of the input vector.
The model architecture of the neural network is shown in Supplementary Fig.~S1.The input is first subdivided into 3 sub-vectors: the $\mathbf{x}_L$ subset of features associated with CDR3 length, the $\mathbf{x}_{A}$ subset of features associated with the CDR3 amino acid composition and the $\mathbf{x}_{VJ}$ subset of features associated with V and J gene usage.
We applied a dense layer individually to $\mathbf{x}_L$ and $\mathbf{x}_{VJ}$. In parallel, we performed an amino acid embedding of $\mathbf{x}_{A}$:  we first reshape the vector to a $2K\times 20$ matrix $\mathbf{A}$ (the set of features associated with amino acid usage is $2\times K\times 20$ long, where $K=25$ the maximum distance from the left and right ends that we encode, and $20$ is the number of amino acids) and apply a linear embedding trough $\mathcal{M}(\mathbf{A})= \mathbf{A\mathbf{M}}$ with $\mathbf{M}$ a $20 \times n$ matrix with $n$ the size of the amino acid encoding. We then flatten the matrix to an array and apply a dense layer. We merged the three transformed subsets into a vector and then applied a dense layer. We finally applied a last dense layer without non-linearity to produce the output value, $\log Q $(see Fig ~S1).\\

The model for paired chains focuses on combining the $\mathbf{x}_L$ and $\mathbf{x}_{VJ}$ inputs of the two chains. First the $\mathbf{x}_L$ and $\mathbf{x}_{VJ}$ inputs within each chain are merged and processed with a dense layer. Subsequently a Batch Normalizing Transform is applied to each encoded vector to enforce a comparable contribution of each chain once the vectors are merged and processed through a dense layer (this last step is skipped in the deep-indep model). A Batch Normalizing Transform \cite{BatchNorm} is a differentiable operator which is normally used to improve performance, speed and stability of a Neural Network.  Given a batch of data, it normalizes the input of a layer such that it will have mean output activation 0 and standard deviation of 1.
In parallel, the amino acid inputs are embedded as described before. Finally all the vectors are merged together and a dense layer without activation outputs the $\log Q$ (see Fig ~S3-4).\\

\section{soNNia model inference}
Given a sample of data sequences $\D=\{\x^i\}_{i=1}^{N_D}$ and a baseline $\G=\{\x'^i\}_{i=1}^{N_G}$ we want to maximize the average log-likelihood:
\begin{equation}
\begin{split}
\mathcal{L}(\theta) & =\langle \log P_{\rm post}^\theta \rangle_\D= \frac{1}{N_D} \sum_{i=1}^{N_D} \log P_{\rm post}^\theta(\x^i)\\
 & =\frac{1}{N_D} \sum_{i=1}^{N_D}[ \log Q^\theta(\x^i) + \log P_{\rm gen}(\x^i)]- \log Z_\theta \\
 & = \langle \log Q^\theta \rangle_\D + \langle \log P_{\rm gen} \rangle_\D - \log \langle Q^\theta \rangle_{\G},
\end{split}
\label{eq:sonialikelihood}
\end{equation}
where $Z_\theta=\<Q^\theta\>_\G=N_G^{-1}\sum_{i=1}^{N_G}Q^\theta(\x'^i)$.
The $P_{\rm gen}$ term in the last equation is parameter independent and can thus be discarded in the inference. When an empirical baseline is used, $P_{\rm gen}$ is replaced by $P_{\rm emp}(\x)=N_G^{-1}\sum_{i=1}^{N_G}\delta_{\x,x'_i}$. \comment{Otherwise, the baseline $\G$ is  sampled from the $P_{\rm gen}$ model, which we learn  from  nonproductive sequences using the IGoR software~\cite{Marcou2018}.\\
}

The above likelihood is implemented in the soNNia inference procedure (linear and non-linear case) with the Keras \cite{chollet2015keras} package. The model is invariant with respect to the transformation $Q(\x)\rightarrow c Q(\x)$ and $Z\rightarrow Z/c $, where c is an arbitrary constant, so we fix dynamically the gauge $Z=1$. We lift this degeneracy by adding the penalty  $\Gamma(\theta)=(Z_\theta-1)^2$,
and minimize $-\mathcal{L}_{sonia}(\theta)+\gamma\Gamma(\theta)$ with $\gamma=1$ as a loss function. \\

In our implementation batch sizes between $10^3-10^4$ sequences produced a reliable inference. L2 and L1 regularization on kernel weights are also applied. Hyperparameters were chosen using a validation dataset of size 10 $\%$ of training data. \comment{The inference converges after around 100 epochs and the network does not overfit (Fig.~S2A). To test the stability of our inference, we evaluated the $P_{\rm post}$ values of generated sequences, based on two models trained on subsets of the initial training data, and show that the $P_{\rm post}$ estimated are highly reproducible between these selection models (Fig S2B).} \\

\comment{The left-right linear SONIA model contains an additional residual gauge, which makes the selection factor invariant with respect to the following transformation:
\begin{equation}
\begin{split}
& q^L_i(a)\rightarrow \lambda_i q^L_i(a) \\
& q^R_j(a)\rightarrow \mu_j q^R_j(a) \\
& q_\ell \rightarrow  q_\ell \prod_{i=\ell+1}^{\ell_\text{max}}\lambda_i \prod_{j=\ell+1}^{\ell_\text{max}} \mu_j 
\end{split}
\end{equation}
where $q^L_i(a)$ and $q^R_j(a)$ are respectively selection factors associated with the usage of  amino acid  $a$ at positions $i,j\in \{1,\dots,\ell_{\rm max}\}$  from the left and the right boundaries of  CDR3 (Fig. 1D),  and $q_\ell$ is the selection factor associated with  CDR3 length $\ell$. The default value of $\ell_\text{max}$ is 25 aa in the left-right model for TCRs. We constrain the gauge by imposing $\sum_a P^L_{i,\mathcal{G}}(a)q^L_i(a)=1$ and $\sum_a P^R_{j,\mathcal{G}}(a)q^R_j(a)=1$ at all positions, similar to \cite{Elhanati2014}. Here, $P^L_{i,\mathcal{G}}(a)$ and $P^R_{j,\mathcal{G}}(a)$ are the marginal probabilities for observing amino acid $a$ at respective positions $i$ (from the left) and $j$ (from the right) of CDR3 in the pre-selected ensemble  $\mathcal{G}$ of sequences.\\
}

To learn the $Q_{\rm VJL}$ model of Fig.~4, we used a linear SONIA model where features $f$ where restricted to $V,J$ and CDR3 length features. One major difference with the approach of Ref.~\cite{EMERSON201314} is that, unlike the likelihood they use, we do not double-count the distribution of length (through $P(L|V)P(L|J)$). However, our results show that that error does not affect model performance substantially.\\

\section{Hierarchy of models in linear SONIA}
The linear SONIA model,
\begin{equation}\label{eq:linearSONIA}
  Q^\theta(\x)=e^{\sum_f \theta_fx_f},
\end{equation}
may be rationalized using the principle of minimum discriminatory information. In this scheme, we look for the distribution $P_{\rm post}$ that is most similar to our prior, described by the baseline set $P_{\rm gen}$ (or empirical set $\G$, replacing $P_{\rm gen}$ by $P_{\rm emp}(\x)=N_G^{-1}\sum_{i=1}^{N_G}\delta_{\x,\x^i}$), but that still reproduces the marginal probabilities in the data. This translates to the minimization of the functional:
\begin{equation}
\begin{split}
\mathcal{F}(P_{\rm post}) &=  {D}_{\rm KL}(P_{\rm \rm post}\Vert P_{\rm gen}) - \eta_0 \Big(\sum_{\x} P_{\rm post}(\x)-1\Big)\\
&- \sum_{f} \theta_f \Big(P_{\rm post}(f) - P_{\rm data}(f)\Big), \label{eq:functional}
\end{split}
\end{equation}
where
\begin{equation}\label{eq:DKL}
  D_{\rm KL}(P_{\rm \rm post}\Vert P_{\rm gen})\doteq \sum_\x P_{\rm \rm post}(\x) \log\frac{P_{\rm \rm post}(\x)}{P_{\rm gen}(\x)}.
  \end{equation}
The second term on the right-hand side imposes the normalization of $P_{\rm post}$ and the last term imposes the constraint that the marginal probabilities of the selected set of features $f$ should match those in the data through the set of Lagrange multipliers $\theta_f$. This scheme reduces to the maximum entropy principle when $\G$ is uniformly distributed.
Minimization of eq.~\ref{eq:functional} results in:
\begin{equation}
P_{\rm post}(\x)=\frac{e^{\sum_{f} \theta_f x_f}}{Z_\theta}P_{\rm gen}(\x),
\label{eq:mindisc}
\end{equation}
where $Z_\theta=e^{1-\eta_0}$, which is equivalent to eq.~\ref{eq:linearSONIA}. Because of the principle of Kullback-Leibler divergence minimization, adding new constraints on the features to the optimization necessary increases $D_{\rm KL}$. This allows us to define a hierarchy of models as we add new constraints.\\

To evaluate the relative contributions of each feature to the difference between CD4 and CD8 TCR, we define different models based on a baseline set $\G$ defined as empirical sequences, with (1) only CDR3 length features; (2) CDR3 length and amino acid features; (3) CDR3 length and VJ features; and (4) all features. We denote the corresponding KL divergences (eq.~\ref{eq:DKL}) $D_{\rm KL}^r(L)$, $D^r_{\rm KL}(A)$, $D^r_{\rm KL}(VJ)$, and $D^r_{\rm KL}({\rm full})$ for each subrepertoire $r=$CD4 or CD8, with $D^r_{\rm KL}({\rm full})\geq D^r_{\rm KL}(A),D^r_{\rm KL}(VJ)\geq D_{\rm KL}^r(L)$. In Fig.~S9 each of these divergences are then combined to get a ``fractional Jensen-Shannon'' divergence $D^f_{\rm JS}=f D_{\rm KL}^{\rm CD4}+(1-f) D_{\rm KL}^{{\rm CD8}}$, where $f$ is the fraction of CD4 cells.\\

\comment{
\section{Transfer Learning}
Training a deep soNNia model (Fig.~1C) for each subset in the analysis of Fig.~4 leads to overfitting issues due to limited data. 
To solve this problem, we can use a mixture technique known as transfer learning (Fig.~S5). Specifically, we first infer a deep soNNia model to characterize selection factors ($\mathcal{Q}^\text{DNN}(x)$) on  unfractionated repertoire data from ref.~\cite{Dean2015}: $P^\text{{emp}}_\post(x)=\mathcal{Q}^\text{DNN}(x)P_\gen(x)$ (eq.~1). We subsequently modulate the distribution by learning an additional linear selection model $\mathcal{Q}^\text{trans}(x)$ for each sub-repertoire,
 \EQ 
 P^\text{trans}_\post(x)= \mathcal{Q}^\text{trans}(x) P^\text{emp}_\post(x)=\mathcal{Q}^\text{trans}(x)\mathcal{Q}^\text{DNN}(x)P_\gen(x).
 \EE
 
If we have enough pooled data,  the deep soNNia model $P^{\rm emp}_\post(x)$ should reproduce the associated empirical distribution of the unfractionated repertoire $P_{\emp}$.  As a result, the first step of this transfer learning algorithm can be replaced by using the empirical distribution $P_{\emp}$ as the common baseline set $\G$, on top of which we can infer a linear selection model with SONIA. The inferred  selection factors  would then reflect deviations from this empirical baseline. Fig.~S5 shows that these two approaches produce very similar selection factors.}

\section{Estimation of information theoretic quantities}
\begin{itemize}

\item {\bf Mutual information} \\
Given two random variables $X$ and $Y$ with joint distribution $p(x,y)$, the mutual information is:
\begin{equation}
I(X,Y)=\sum_{x,y}p(x,y)\log{\frac{p(x,y)}{p(x)p(y)}},
\label{eq.mutinfo}
\end{equation}
and $P(x)$ and $P(y)$  are the respective marginal distributions of $p(x,y)$. 
$I(X,Y)$ can be naively estimated from data through the empirical histogram $(x,y)$. The estimated mutual information $\hat{I}$ on a finite sample of data is affected by a systematic error \cite{Steuer2002}.  We estimated the finite sample systematic error $I_0(X,Y)$ by destroying the correlations in the data through randomization. We implemented the randomization by mismatching CDR3-length, V and J assignment within the set. This mismatching procedure leads to the same marginals, $P(V)$ or $P(J)$, but destroys correlations, $P(V,J)-P(V)P(J)\simeq 0$.\\

 \comment{\item {\bf Jensen-Shannon divergence} \\
To quantify differential selection, we evaluate Jensen-Shannon divergence $D_{\rm JS}(P_\post,P'_\post)$ between pairs  $(r,r')$ of sub-repertoires, $P_\post^{r}$ and $P_\post^{r'}$,

\beqn
\nonumber D_{\rm JS}(P_\post,P'_\post)&=&\frac{1}{2}D_{\rm KL}(P_\post \Vert (P_\post+P'_\post)/2)+\frac{1}{2}D_{\rm KL}(P'_\post\Vert (P_\post+P'_\post)/2)\\
&=&\frac{1}{2}\Big\langle \log_2 \frac{2\Q^{r}}{\Q^{r} +\Q^{r'}} \Big\rangle_{r}
+ \frac{1}{2} \Big\langle \log_2 \frac{2\Q^{r'}}{\Q^{r} +\Q^{r'}} \Big\rangle_{r'}
\eeqn
where $\langle\cdot\rangle_{r} $ denotes averages over $P_\post^r$.\\

\item {\bf Entropy of  paired receptor repertoires}\\
To quantify diversity of immune receptors associated with paired chains, we estimated the entropy 
\beq
H(P_{\post})=-\langle \log_2 P_{\post}\rangle_{P_{\post}}=-\langle \mathcal{Q} \log_2 P_{\post}\rangle_{P_{\gen}}
\eeq
of the paired chain models by sampling $10^6$ sequences from the generation model $P_{\gen}$ (Table S1). For comparison, we evaluated the entropy of single chain repertoires, using models inferred for each chain separately. We also evaluated the entropy of V and J gene features using the observed marginal probabilities of these features in the data. For example, the entropy associated with  V-genes in the heavy chain repertoire can be calculated as $H(P_{\post}^{V_H})=-\sum_i P(V^i_H) \log_2 P(V^i_H) $, where   $P(V^i_H) $ is the marginal probability for the  $i^{th}$ V-gene  in a heavy chain (H) dataset.   \\
}
\end{itemize}

\comment{Errors in estimating Entropy $H$ (Table S1),  the Kullback-Leibler divergences $D_{\rm KL}$ (Fig.~2) and Jensen-Shannon divergences $D_{\rm JS}(P_\post,P'_\post)$ (Figs.~4,~S7) are evaluated by computing the standard deviation of the above quantities using subsampled datasets of size one fifth of the original data. Here we assume that $P_\post (r)=\frac{1}{Z} P_\gen (r) \Q(r) $ or $P_\post(r)= \frac{1}{Z}  P_\emp(r) \Q(r)$, with $P_\gen$ or $P_\emp$ as baselines, respectively (eq.~1).\\
}

\small{

}

\clearpage{}
\newpage{}
\renewcommand{\thetable}{S\arabic{table}}

\begin{table*}[h!]
\small
\centering
\begin{tabular}{c|ccccccc}
{\bf cell type}&\multicolumn{7}{c}{\multirow{1}{*}{{\bf entropy [bits]}}}\\
&$H(P^{HL}_{\rm post})$&$H(P^{H}_{\rm post})$&$H(P^{L}_{\rm post})$ & $H(V_{\rm H})$ & $H(J_{\rm H})$ & $H(V_{\rm L})$ & $H(J_{\rm L})$  \\ 
\hline
TCR $\alpha \beta$  & $54.5\pm0.1$ & $31.4\pm 0.1$&$22.9\pm 0.1$& $4.803 \pm 0.004$& $3.404 \pm 0.001$& $4.906\pm 0.002$& $5.381 \pm 0.001$ \\  
Ig H$\lambda$ & $57.0 \pm 0.1$ & $44.5\pm 0.4$&$14.0\pm 0.1$& $4.67\pm 0.02$& $1.98\pm0.02$& $3.81\pm0.01$& $1.40\pm0.01$ \\  
Ig H$\kappa$  & $58 \pm 1$ & $44.5\pm 0.4$&$12.9\pm 0.1$& $4.74\pm0.01$& $2.04\pm0.01$& $3.64\pm0.02$& $2.24\pm0.01$  \\  
\end{tabular}
\caption{\comment{Entropy contribution from different receptor features  in the paired-chain selection models. These entropy values are estimated based on the amino acid content of receptors' CDR3 and their  V-,J- gene usages.}}
\end{table*}

\renewcommand{\thefigure}{S\arabic{figure}}
\setcounter{figure}{0}

\begin{figure*}
\centering
\includegraphics[width=0.75\linewidth]{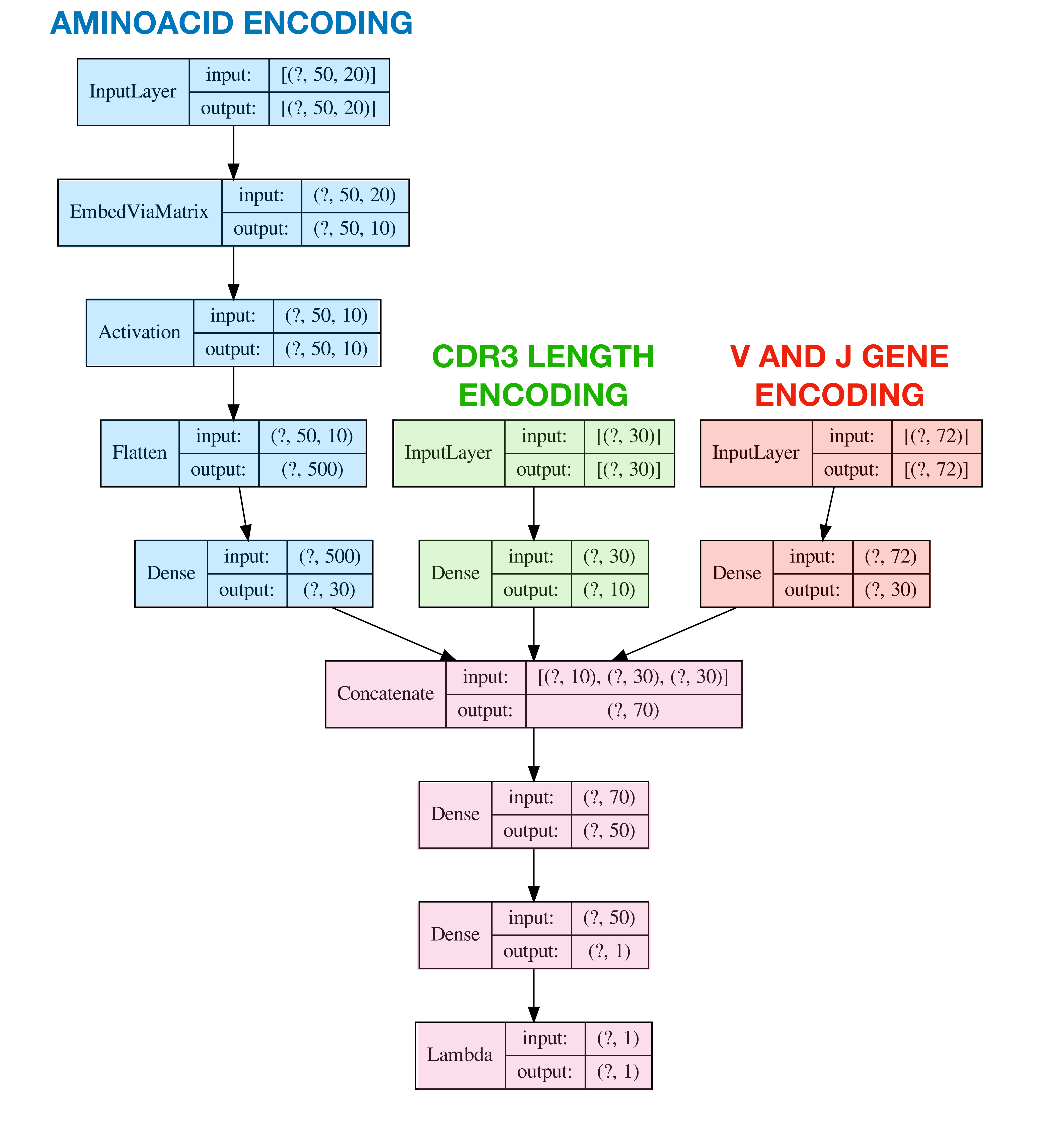}
\caption{Neural network structure of the deep soNNia model for the single chain case. There are three inputs, from left to right: first the encoded aminoacid composition of the CDR3 using the left-right encoding scheme, then the length of the CDR3, finally the independent V and J gene usage information. The aminoacid input is encoded using an embedding layer, called EmbedViaMatrix  and then processed by a tanh non-linearity, called Activation layer. The Flatten layer turns the encoded matrix in the corresponding flattened array where each row of the matrix is concatenated to the successive one. A dense layer is then applied to reduce its dimensionality. The other two inputs are also processed through  a dense feed-forward layer to reduce their corresponding dimensionality. The three groups of encoded inputs are then concatenated and two dense feed forward layers are applied to output $\log Q$. Finally $\log Q$ is clipped to avoid diverging values using the Lambda layer.}
\end{figure*}

\begin{figure*}
\centering
\includegraphics[width=0.75\linewidth]{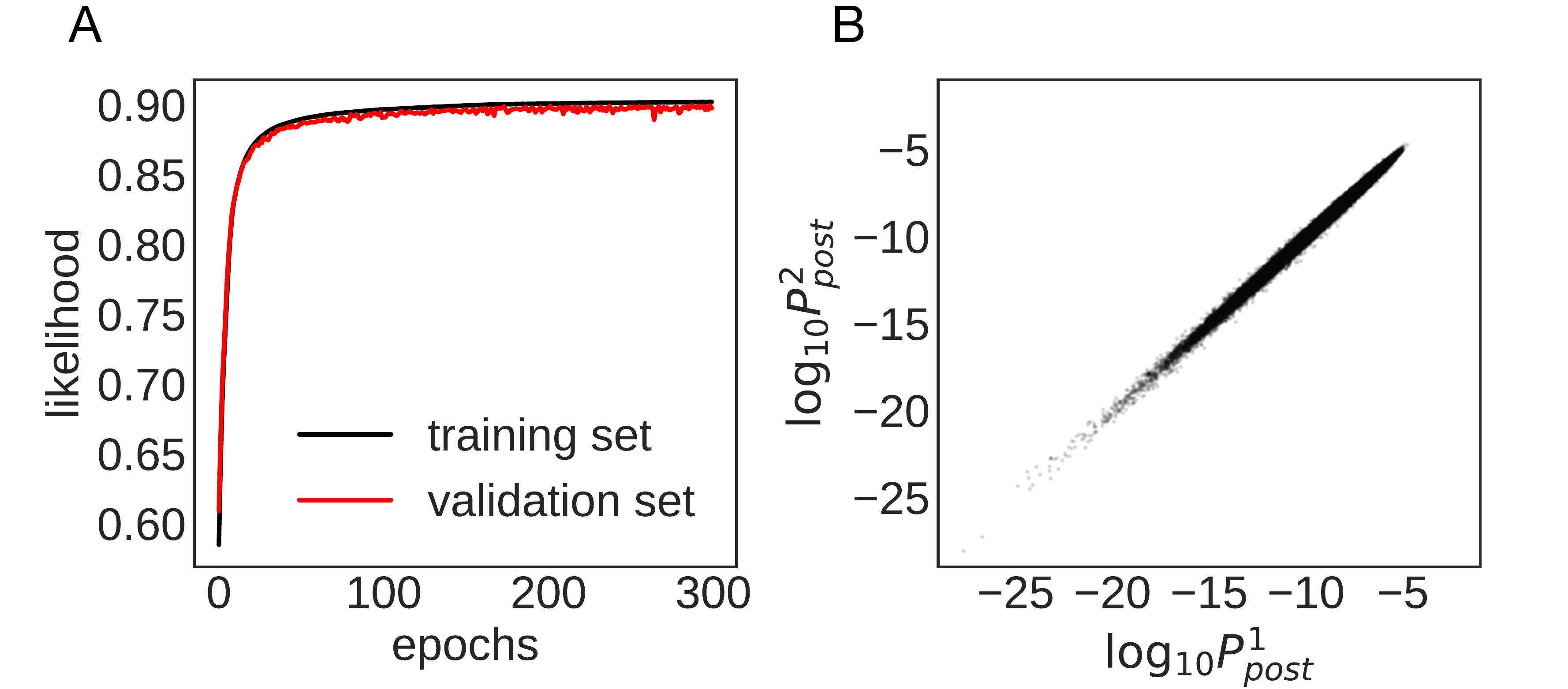}
\caption{\comment{{\bf (A)} Convergence of the training and validation likelihoods as a function of training epochs for the soNNia model shown in Fig.~2B. {\bf (B)} Comparison between two soNNia models trained on independent datasets obtained by splitting the training set of $10^7$ TCR$\beta$ sequences pooled from repertoires of 743 individuals of ref.~\cite{Emerson2017} (Fig. 2) in two equal parts.}}
\end{figure*}

\begin{figure*}
\centering
\includegraphics[width=.95\linewidth]{figs/FIGS5.pdf}
\caption{Neural Network structure of the deep-indep model for paired chains. See Fig.~S1 for details on  each layer.}
\end{figure*}

\begin{figure*}
\centering
\includegraphics[width=.95\linewidth]{figs/FIGS4.pdf}
\caption{Neural Network structure of the deep-joint model for paired chains. See Fig.~S1 for details on  each layer.
 }
\end{figure*}

\begin{figure*}
\centering
\includegraphics[width=\linewidth]{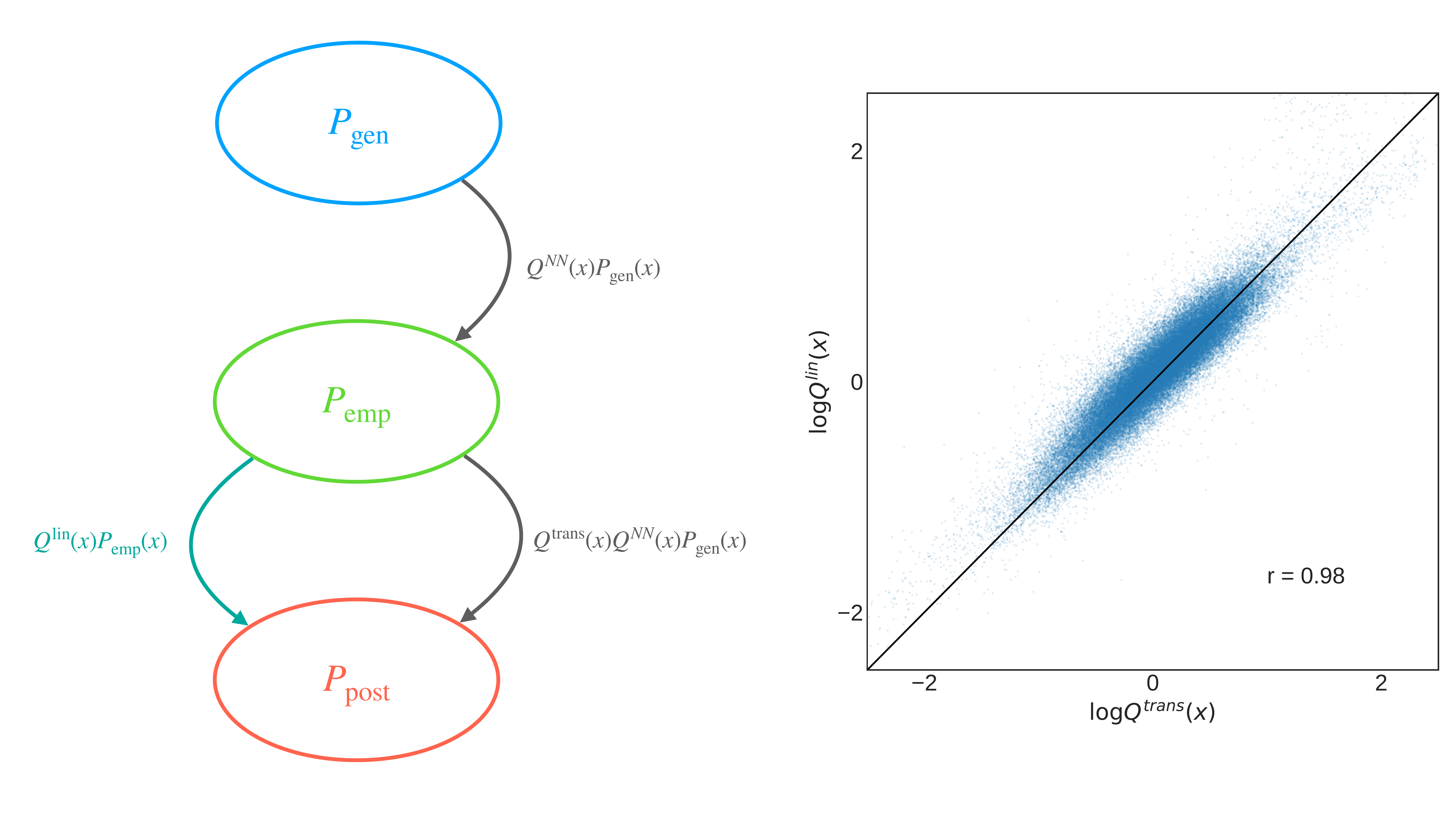}
\caption{Transfer learning consists of a 2-step inference (left): in the first step we infer a deep neural network on a bigger data set $\G$, and in the second second step we re-infer a subsection of the network, or an additional layer on a smaller dataset, which is the real target. In our specific application, the big data $\G$ is the unfractioned repertoire from ref.~\cite{Dean2015} ($P_{\rm emp}(\x)=N_G^{-1}\sum_{i=1}^{N_G}\delta_{\x,x'_i}$), and the final targets are the sub-repertoires harvested from different tissues~\cite{Seay2016}. We can infer a deep selection model soNNia to characterize well the unfractioned repertoire $P_{\rm emp}$, and then learn a functional selection model for each sub-repertoire with an additional linear layer in the neural network. This procedure is equivalent to using $P_{\rm emp}$ as the baseline distribution in the inference of a linear selection model, as it can be seen by the high correlation between selection factors inferred with the two different methodologies (right). 
}
\end{figure*}

\begin{figure*}
\centering
\includegraphics[width=\linewidth]{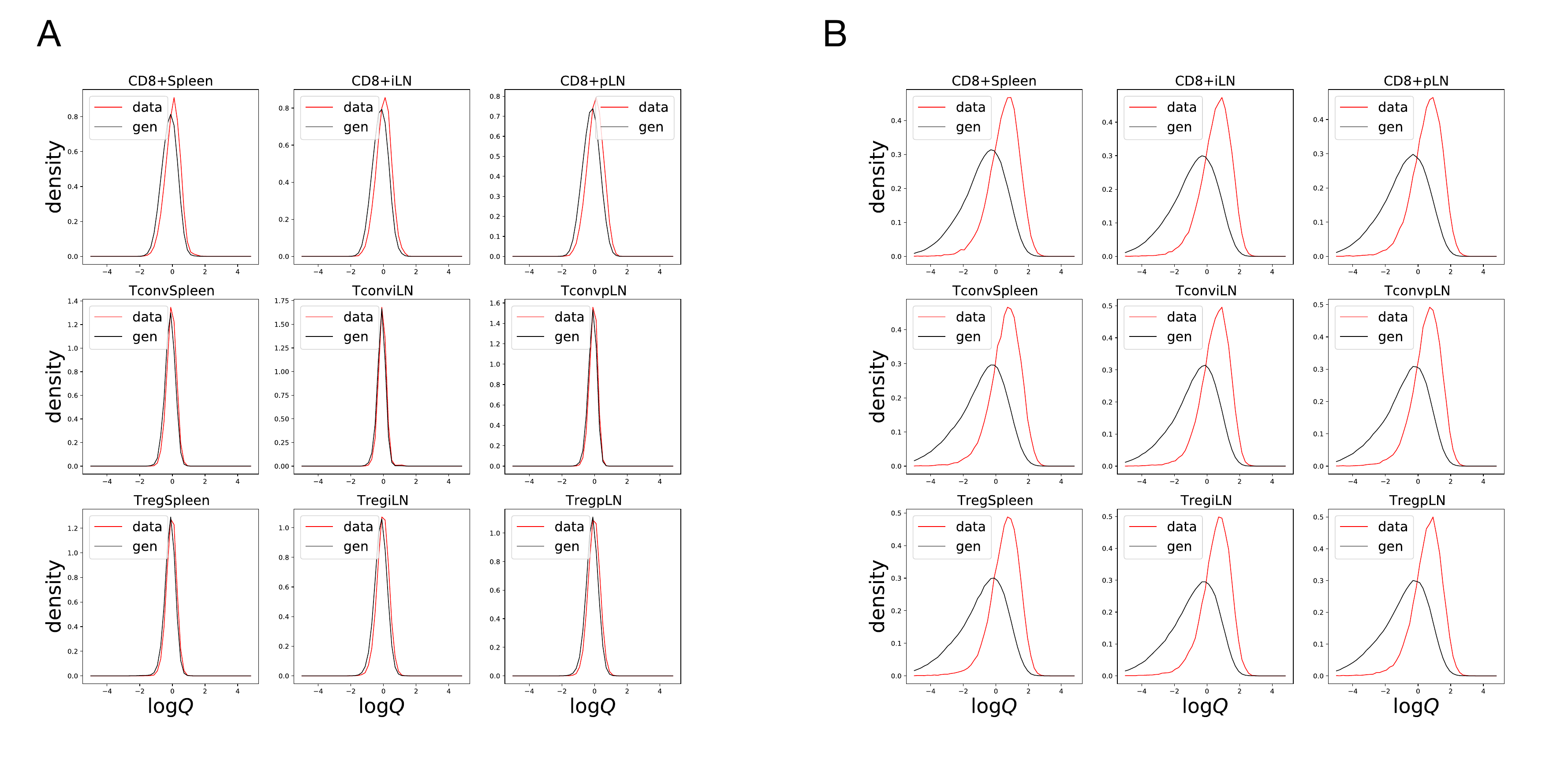}
\caption{{\bf(A)} Distribution of $\log\mathcal{Q}$ of inferred models starting from an empirical baseline $\G$, and {\bf(B)} the distribution of $\log\mathcal{Q}$ of inferred models starting  from the $P_{\rm gen}$ model as a baseline.}
\end{figure*}
\begin{figure*}
\centering
\includegraphics[width=\linewidth]{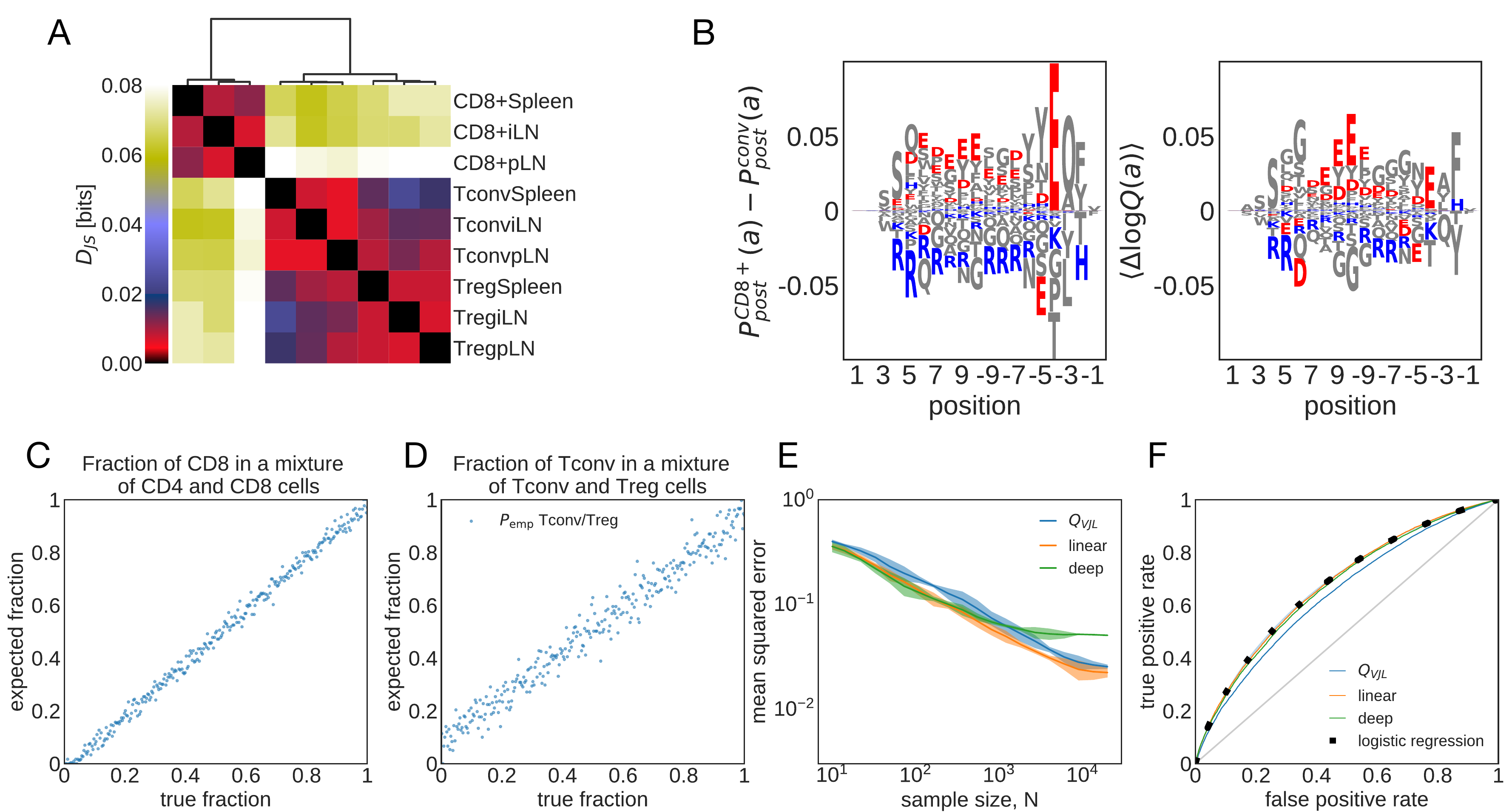}
\caption{Analogous to Fig 4 in main text but with $P_{\rm gen}$ as null model. {\bf (A)} Jensen-Shannon divergences ($D_{JS}$, see eq.~8 in Methods) computed from models trained on different sub-repertoires. \comment{{\bf (B)} 
Difference in the marginal probability  for amino acid composition along CDR3 between CD8$^+$ and CD4$^+$ Tconv (left) and the expected difference in the corresponding log-selection factors for amino acid usage (right)  are shown. The negatively charged amino acids (Aspartate, D, and Glutamate, E) and the positively charged amino acids (Lysine, K, and Arginine, R) are indicated in red and blue, respectively. Other amino acids are shown in gray.} {\bf(C)} Maximum-likelihood inference of the fraction of CD8$^+$ TCRs in mixed repertoires of Tconv and CD8$^+$ cells from spleen (eq.~4). Each repertoires comprises $5\times 10^3$ unique TCR. {\bf(D)} Same as (C) but for a mixture of Tconv and Treg TCR. {\bf(E)} Mean squared error of the inferred sample fraction from (C) as a function of sample size $N$, averaged over all fractions, using models of increasing complexity: ``$Q_{VJL}$" is a linear model with only features for CDR3 length and VJ usage, ``linear" is linear SONIA model, `deep' is the full soNNia model (see Fig.~1C).
  {\bf(F)} Receiving-Operating Curve (ROC) for classifying individual sequences as coming from CD8$^+$ cells or from CD4$^+$ conventional T cells from spleen, using the log-likelihood ratios.
Curves are generated by varying the threshold in eq.~5. The accuracy of the classifier is compared to a traditional logistic classifier inferred on the same set of features as our selection models. The training set for the logistic classifier has $N=3 \times 10^5$ Tconv CD4$^+$, $N= 8.7 \times 10^4$ CD8$^+$ TCRs, and the test set has $N= 2 \times 10^4$ CD4, $N= 2 \times 10^4$ CD8$^+$ TCR sequences.}
\end{figure*}

\begin{figure*}
\centering
\includegraphics[width=\linewidth]{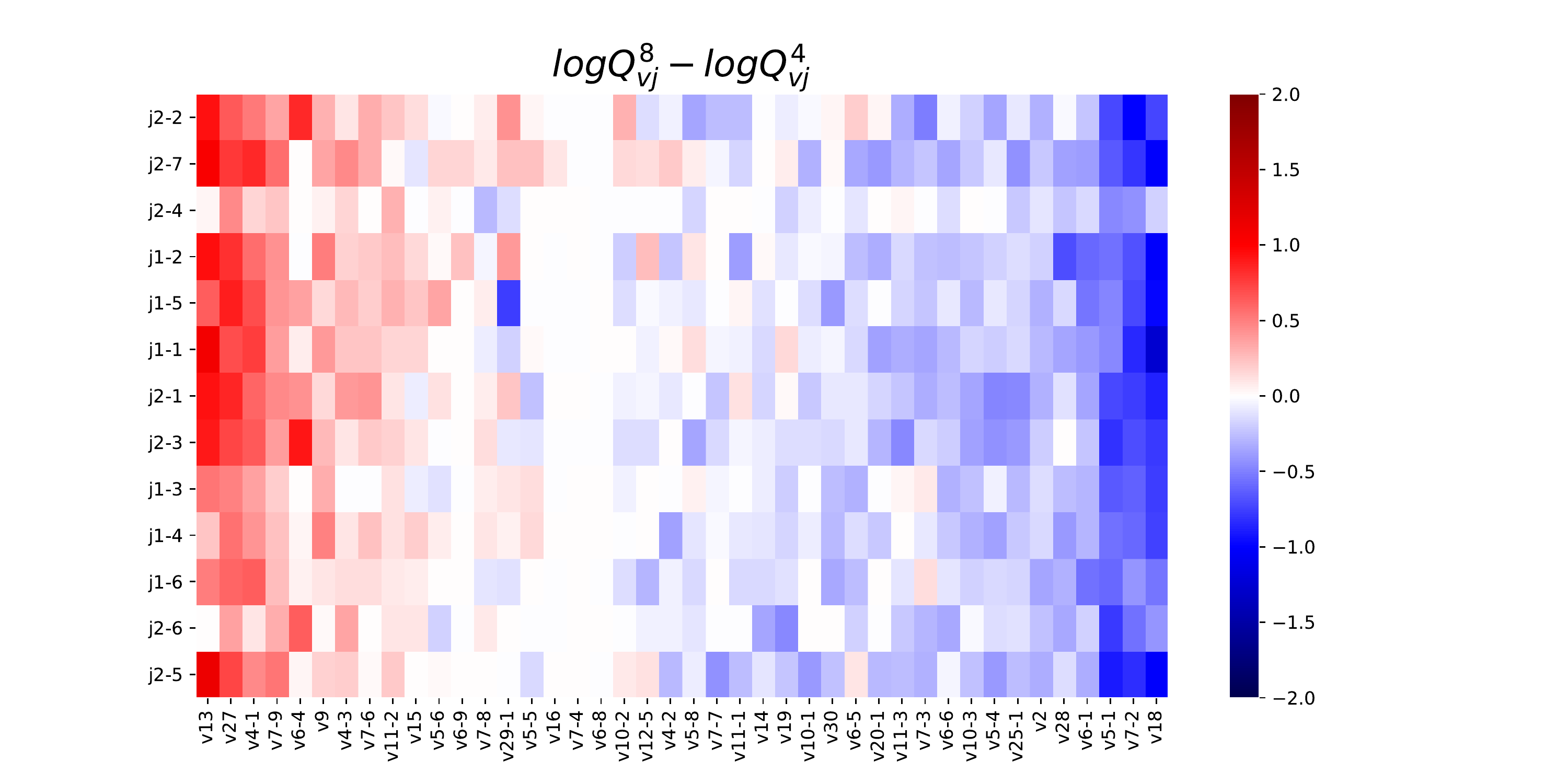}
\caption{Differential selection on V and J gene usage between CD4$^+$ and CD8$^+$ models {\color{black} inferred on top of $P_{\rm emp}$ as baseline distribution.} 
}
\end{figure*}

\begin{figure*}[t!]
\centering
\includegraphics[width=0.6\linewidth]{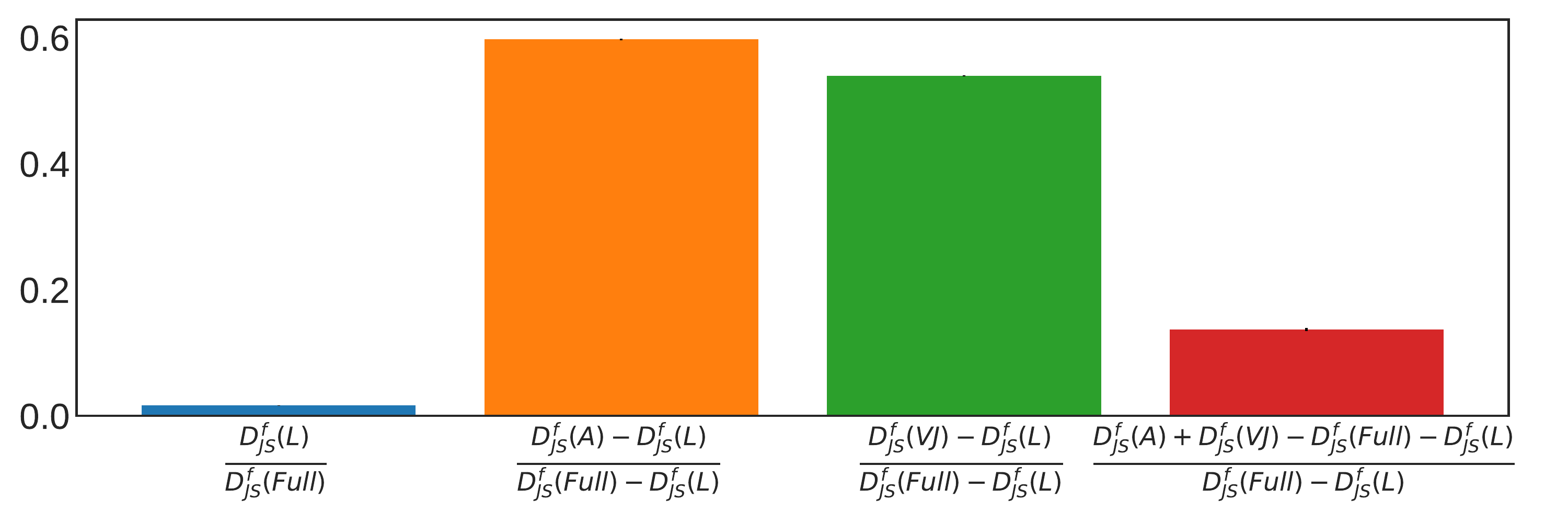}
\caption{Decomposition of contribution from different features to the fractional Jensen Shannon divergence between the CD4 and CD8 subpertoire statistics, $D^f_{\rm JS}(L)\leq D_{\rm JS}^f(A),D_{\rm JS}^f(VJ)\leq D_{\rm JS}^f({\rm full})$. The blue bar is the contribution of CDR3 length; orange and green bars are the relative contributions from the amino-acid composition and VJ usage, respectively. Red bar is the fraction that's redundant between VJ and amino acid usage. Contributions are balanced between amino acid and VJ usage, with moderate redundancy between the two.}
\end{figure*}

\begin{figure*}
\centering
\includegraphics[width=\linewidth]{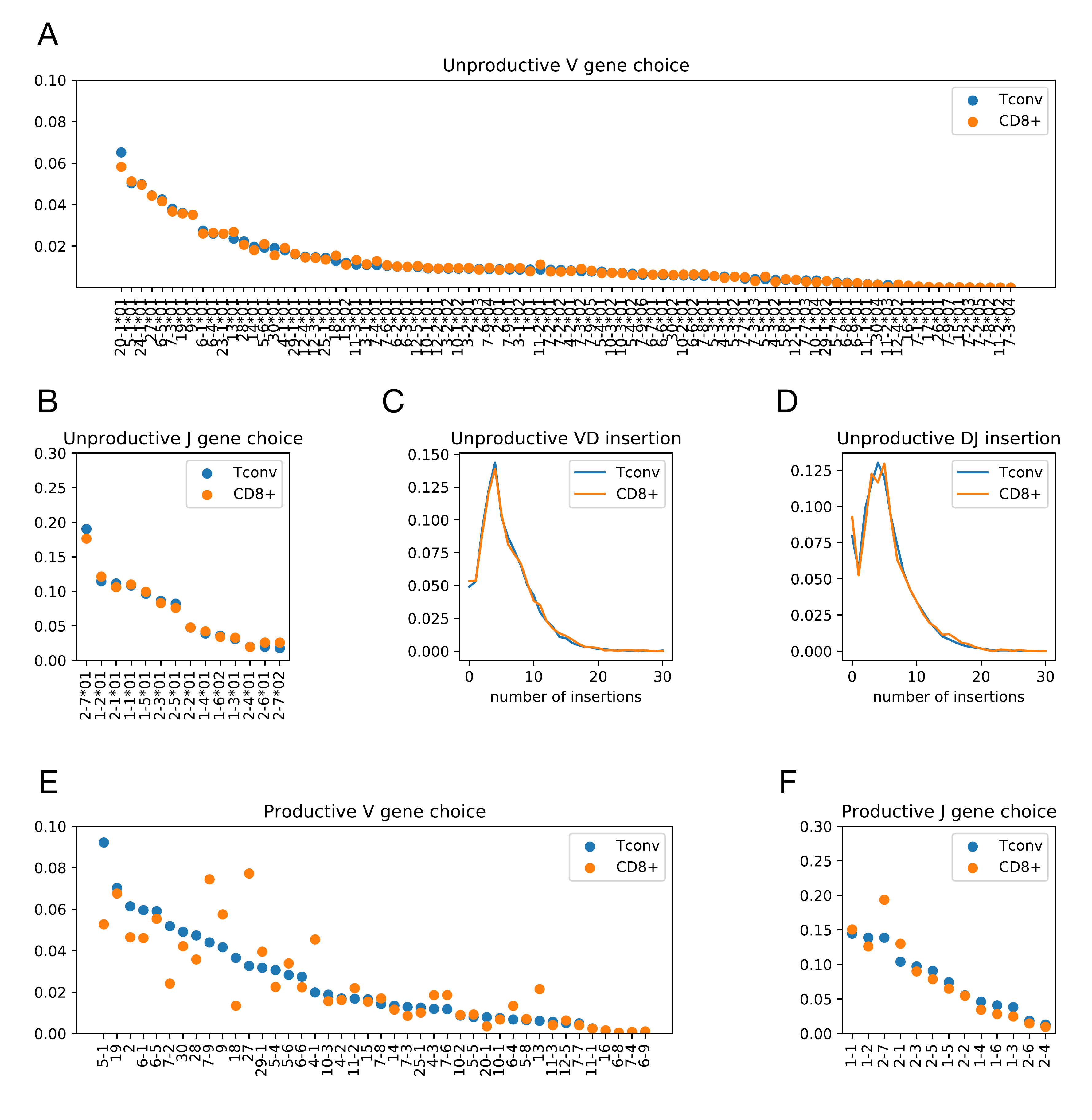}
\caption{\color{black}{\bf (A-D)} Parameters of generation models between CD4$^+$ and CD8$^+$ T cells do not differ significantly. {\bf(E,F)} V and J gene usages in productive CD4$^+$ Tconv and CD8$^+$ T cells, for comparison to (A,B). }
\end{figure*}

\begin{figure*}
\centering
\includegraphics[width=0.7\linewidth]{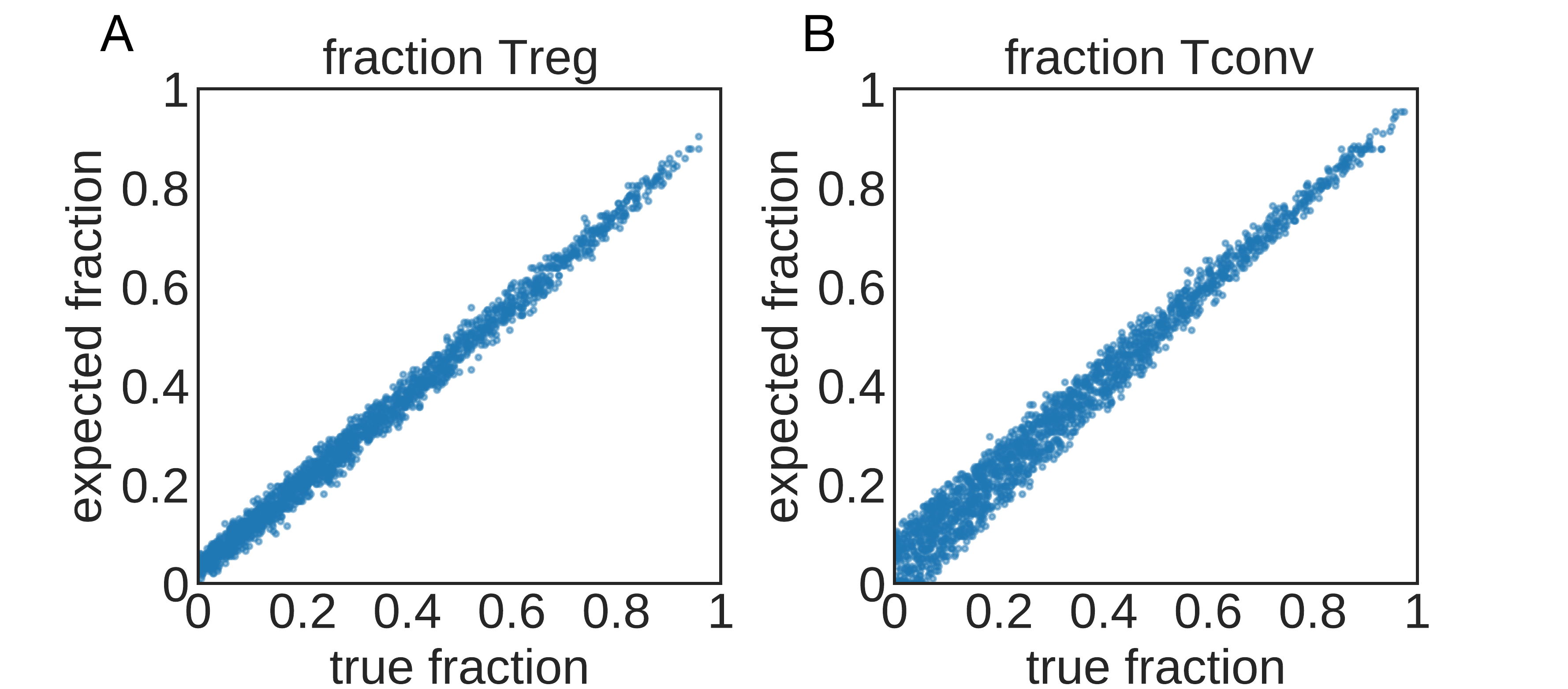}
\caption{Joint Inference of fraction of {\bf (A)} Treg and {\bf (B)} CD4$^+$ Tconv cells, belonging to different subclasses in a mixture of 3 repertoires: CD8$^+$,CD4$^+$ Tconv, and CD4$^+$ Treg cells. We  optimized the likelihood $L(f_1,f_2)=\sum_i( f_1 Q_{\rm conv}(x_i)+ f_2 Q_{\rm reg}(x_i) +(1-f_1-f_2)Q_{CD8}(x_i))$ to infer jointly the two fractions $f_1$ and $f_2$ in a chosen mixture of  $3\times 10^4$ TCRs $x_i$, built by combining repertoires of purified subsets harvested from spleen~\cite{Seay2016}. Each point corresponds to a mixture with $f_1$ and $f_2$ sampled uniformly 2000 times in the simplex $f_1\geq 0$, $f_2\geq 0$, $f_1+f_2\leq 1$.}
\end{figure*}
\begin{figure*}
\centering
\includegraphics[width=0.9\linewidth]{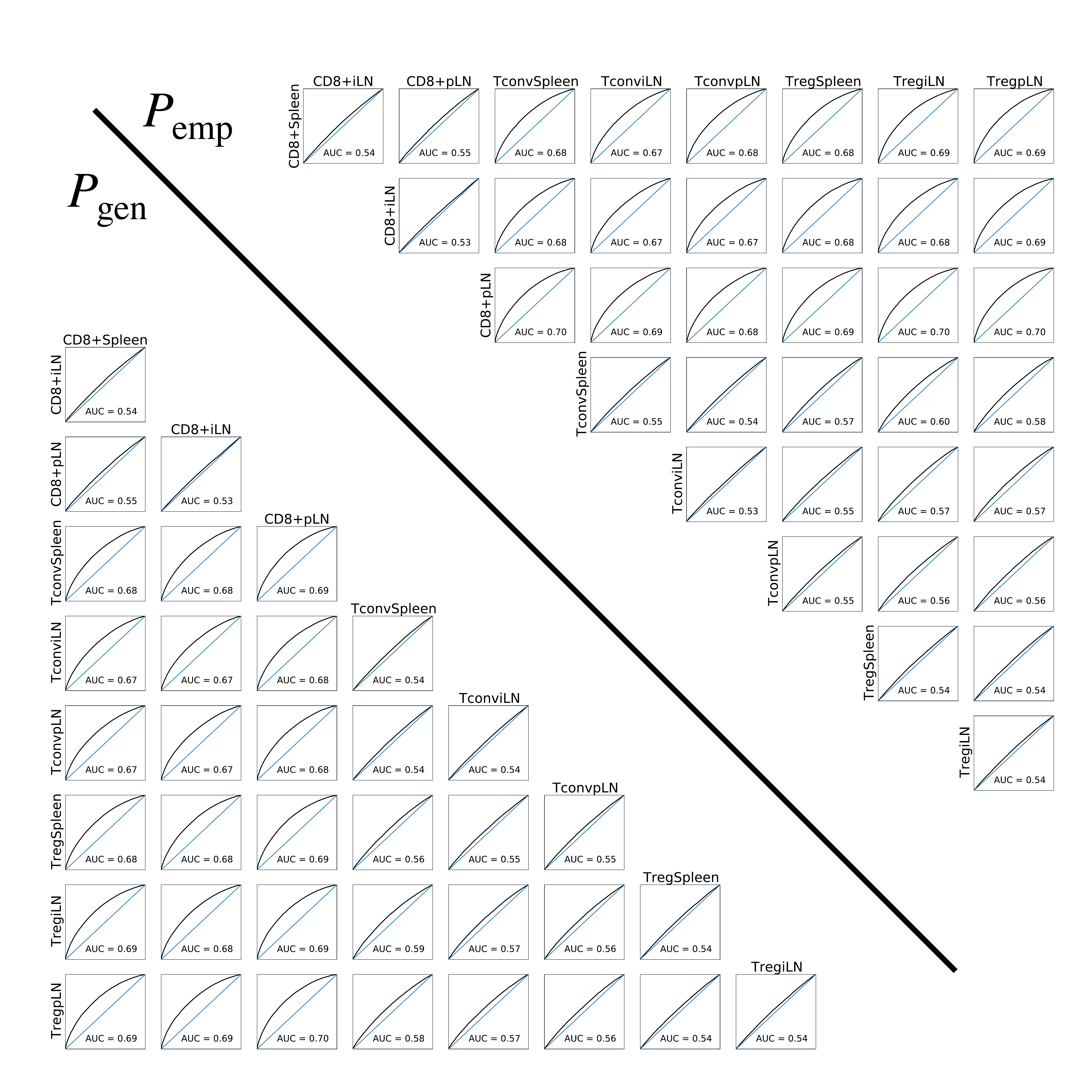}
\caption{ROC curve between all subsets based on the log ratio $R(x)$ defined on main text, where the selection factors are inferred starting from the empirical baseline $\G$ ($P_{\rm emp}(\x)=N_G^{-1}\sum_{i=1}^{N_G}\delta_{\x,x'_i}$, above diagonal) or $P_{\rm gen}$ (below diagonal).}
\end{figure*}